\documentclass[twocolumn]{aastex631}
\usepackage{graphicx} 
\usepackage{amsmath} 
\usepackage{amssymb} 
\usepackage{natbib}
\usepackage{appendix}
\usepackage{svg}
\usepackage{mathtools}
\DeclarePairedDelimiter{\norm}{||}{||}

\begin{document}

\title{Resonance Capture of a Test Particle by an Eccentric Planet in the Presence of Externally-Driven Apsidal Precession}
\author[0000-0002-0896-7393]{JT Laune}
\affiliation{Department of Astronomy, Cornell Center for Astrophysics and Planetary Science, Cornell University, Ithaca, NY 14853, USA}
\author[0000-0002-1934-6250]{Dong Lai}
\affiliation{Department of Astronomy, Cornell Center for Astrophysics and Planetary Science, Cornell University, Ithaca, NY 14853, USA}
\affiliation{Tsung-Dao Lee Institute, Shanghai Jiao Tong University, 200240 Shanghai, China}

\begin{abstract}
Planets undergoing convergent migration can be captured into mean-motion resonance (MMR), in which the planets' periods are related by integer ratios. 
The dynamics of MMR are typically considered in
isolation, including only the forces between the planets and with the central star.  
However, the planets are often subjected to external forces that induce apsidal precessions, which may split the MMR into two sub-resonances and give rise to chaotic motion due to resonance overlap.
In this study, we investigate how such externally induced differential apsidal precession affects capture into first-order $j:j+1$ MMRs. 
We study the restricted three-body problem for a test particle outside of an eccentric planet, with the planet undergoing outward migration.
We find that capture can be sensitive to the differential apsidal precession, even when the precession rate is much smaller than the resonance overlap criterion.
We identify two critical precession frequencies -- related to resonance overlap and secular apsidal resonance -- around which capture is disrupted even for very small planet eccentricity.
Our results may help clarify the capture process of resonant trans-Neptunian objects by outward-migrating Neptune in early Solar System.
\end{abstract}

\section{Introduction}
Mean-motion resonances (MMRs) occur when two satellites/planets orbit a common primary with orbital periods near a commensurate ratio (e.g., $1:2$, $1:3$, $2:3$, etc.).
Planets undergoing convergent migration---with the period ratio ($<1$) increasing in time---can become locked into MMRs~\citep[e.g.,][]{peale1976ARA&A..14..215P}.
This resonant capture can occur in a variety of astrophysical contexts, such as protoplanetary disks with embedded planets, Saturn’s moon system, AGN disks with embedded stellar-mass black holes, and the early Solar System during Neptune’s outward migration.

The first-order $j:j+1$ MMR occurs at the period ratio
\begin{align}
\frac{P_{\rm p}}{P} \simeq \frac{j}{j+1},
\end{align}
where $P_{\rm p}$, $P$ are the periods of the inner and outer bodies (``planets''), respectively, and $j\geq1$ is an integer.
``First-order'' refers to the strength of the resonant interaction between the planets in terms of their eccentricities.
The resonance angles are 
\begin{align}\label{eq:resangles}
    \theta_{\rm p} &= (j+1)\lambda-j\lambda_{\rm p}-\varpi_{\rm p},\\
    \theta &= (j+1)\lambda-j\lambda_{\rm p}-\varpi,\label{eq:resangles1}
\end{align}
where $\lambda_{\rm p},\lambda$ and $\varpi_{\rm p},\varpi$ are the satellites' mean longitudes and longitudes of perihelion.
If the resonant interaction terms ($\propto\cos\theta$ and $\cos\theta_{\rm p}$) dominate the dynamics and other forcing terms on the planets are neglected, the two angles $\theta$ and $\theta_{\rm p}$ can be combined into a single mixed resonance angle \citep[see, e.g.,][]{henrardReducingTransformationApocentric1986,wisdomCanonicalSolutionTwo1986}, the two ``sub-resonances'' (associated with $\theta$ and $\theta_{\rm p}$) merge into one, and the system is integrable.
However, often the planets can be subjected to ``external'' forcings other than the gravity from the central primary, and they experience externally-driven apsidal precession
\begin{equation}\label{eq:oms}
    \omega \equiv \left(\frac{\mathrm{d}\varpi}{\mathrm{d}t}\right)_{\rm ext}\quad\text{and}\quad
    \omega_{\rm p} \equiv \left(\frac{\mathrm{d}\varpi_{\rm p}}{\mathrm{d}t}\right)_{\rm ext}.\\
\end{equation}
In the presence of differential apsidal precession $\Delta\omega_{\rm ext}=\omega_{\rm p}-\omega$, the two sub-resonances are separated in the frequency space by $\Delta\omega_{\rm ext}$.
In this case, the problem may become nonintegrable, and it is possible for resonance overlap to occur.
This may lead to chaos and could influence the capture process.

MMR capture is especially relevant to the history of the resonant population of Kuiper Belt objects (or trans-Neptunian objects, TNOs).
Neptune is thought to have migrated outward to its current-day orbit early in the history of the Solar System \citep{fernandezDynamicalAspectsAccretion1984,malhotraOriginPlutosPeculiar1993,malhotraOriginPlutoOrbit1995,gomesPlanetaryMigrationPlutino2000}.
During its migration, it swept up TNOs from the cold classical belt into its current  outer MMR state..
The 2:3 and 1:2 MMRs with Neptune are particularly interesting, as their dynamical structure charts the early history of the Kuiper Belt \citep{murray-clayUSINGKUIPERBELT2011} and explains the orbit of Pluto.
While most studies of resonance capture have included Jupiter, Saturn, and Uranus in their simulations~\citep[e.g.,][]{gomes2003Icar..161..404G}, up until now the specific contribution of differential apsidal precession on the capture process remains unclear.

Some authors have previously studied the effects of $\Delta\omega_{\rm ext}$ on first-order MMRs.
\cite{murrayEffectsDiskInduced2022} consider $\Delta\omega_{\rm ext}$ induced by a gas disk, and examine how $\Delta\omega_{\rm ext}$ influences the equilibrium behavior of two comparable-mass planets locked into a MMR.
After the disk dissipates and $\Delta\omega_{\rm ext}$ vanishes, significant changes in the resonance architecture occur, and the planets may be ejected from the resonance.
\citet{elmoutamidCouplingCorotationLindblad2014} investigate the coupling between the $\theta$ (``Lindblad'') and $\theta_{\rm p}$ (``corotation'') resonances.
They diagnose chaos in three different regimes of differential apsidal precession, $\Delta\omega_{\rm ext}=0$, $\Delta\omega_{\rm ext}$ near the resonance width, and $\Delta\omega_{\rm ext}$ much larger than the resonance width.
The system is chaotic in the intermediate case but regular for the two extremes.
\citet{elmoutamidDerivationCaptureProbabilities2017} study the capture into the $\theta_{\rm p}$ resonance in the presence of large differential apsidal precession when the two sub-resonances are well-separated. 
It is not clear to what extent such ``clean'' $\theta_{\rm p}$-resonance operates in realistic situations.

In this paper, we study the general dynamics of MMR capture in the eccentric restricted three-body problem (i.e., one of the planets is a test mass) for $\Delta\omega_{\rm ext} \neq 0$.
In Section~\ref{sec:methods}, we describe our problem setup and numerical methods, and in Section~\ref{sec:resdyn}, we examine resonance overlap.
In Section~\ref{sec:Migration}, we allow the planet to migrate, include its secular forcing on the test particle, and study the resulting capture outcomes.
We discuss our results in Section~\ref{sec:discussion} and conclude in Section~\ref{sec:conclusion}.

\section{Problem Setup and Method}\label{sec:methods}
We consider a test particle and a planet, each on orbits around a central star (mass $M$) with semimajor axes (SMAs) $a$ and $a_{\rm p}$ ($<a$), mean motions $n$ and $n_{\rm p}$, eccentricities $e$ and $e_{\rm p}$, and longitudes of perihelion $\varpi$ and $\varpi_{\rm p}$, respectively.
Let the planet have mass $m_{\rm p}$ such that the mass ratio $\mu_{\rm p} \equiv m_{\rm p}/M\ll 1$.
We assume both the test particle and planet are undergoing secular apsidal precession induced by an external source (such as another planet or a disk) at frequencies $\omega$ and $\omega_{\rm p}$ (equation~\ref{eq:oms}).
For simplicity, we assume $\omega$ and $\omega_{\rm p}$ are constant parameters and $\varpi_{\rm p}=0$ at $t=0$, i.e. $\varpi_{\rm p}=\omega_{\rm p} t$.

The outer $j:j+1$ MMR of the test particle with the planet occurs where $n/n_{\rm p}\simeq j/(j+1)$.
The ratio of SMAs, $\alpha=a_{\rm p}/a$, is $\alpha_0\equiv(j/(j+1))^{2/3}$ at exact resonance for $\Delta\omega_{\rm ext}=0$.
In order to simulate resonance capture, we initialize $a_{\rm p}/a<\alpha_0$ and gradually increase $a_{\rm p}$ over time (so that $\dot n_{\rm p}<0$), allowing the resonance location to sweep across the test particle's orbit.
The mean motion of the planet is initialized with $n_{\rm p}=n_{\rm p,0}$ and decreases as
\begin{equation}\label{eq:npdot}
    \frac{n_{\rm p}}{\dot n_{\rm p}} = -\tau_{\rm m},
\end{equation}
where $\tau_m$ is the migration timescale, and $n_{\rm p}\tau_{\rm m}\gg 1$ ensures slow migration. 
Unless explicitly stated otherwise, we set $\tau_m=10^{6}P_{\rm p,0}$, where $P_{\rm p,0}=2\pi/n_{\rm p,0}$ is the planet's initial orbital period.

Near the resonance, the Hamiltonian for the test particle is dominated by resonant terms with angles given by equations~(\ref{eq:resangles}--\ref{eq:resangles1}). Note that
\begin{equation}
    \lambda_{\rm p} = \int_0^t n_{\rm p} dt',\quad
    \dot\lambda_{\rm p} = n_{\rm p},\quad
    \ddot\lambda_{\rm p} = -\frac{n_{\rm p}}{\tau_{\rm m}}.
\end{equation}
The Hamiltonian for the test mass is given by 
\begin{align}\label{eq:H}
    H &= -\frac{GM}{2a}
    -\frac12\sqrt{GMa}~e^2\omega \nonumber\\
    &\qquad+\frac{Gm_{\rm p}}{a_{\rm p}}\beta_{\rm p} e_{\rm p}\cos\theta_{\rm p}-\frac{Gm_{\rm p}}{a_{\rm p}}\beta e\cos\theta,
\end{align}
where the coefficients $\beta_{\rm p},\beta>0$ are functions of the Laplace coefficients \citep{murraySolarSystemDynamics2000},
\begin{align}
    \beta_{\rm p} &= -\frac12\alpha [-2(j+1)-\alpha D]b_{1/2}^{(j+1)}(\alpha), \\
    \beta &= \frac12\alpha[-1+2(j+1)+\alpha D]b_{1/2}^{(j)}(\alpha).
\end{align}
For simplicity, we assume they are evaluated at $\alpha=\alpha_0$ and remain constant throughout the integration.
For $j=2$ (i.e., the 2:3 resonance), $\beta_{\rm p} \approx 1.6$ and $\beta \approx 1.9$.
The action-angle variables are
\begin{align}
    \Lambda=\sqrt{GMa}&\longleftrightarrow \lambda, \\
    \Gamma=\Lambda(1-\sqrt{1-e^2})\simeq \frac12\Lambda e^2 &\longleftrightarrow -\varpi,
\end{align}
where the approximation for $\Gamma$ is valid for small $e$.

In addition to the resonant interactions described by the Hamiltonian $H$ (equation~\ref{eq:H}), the test particle experiences secular forcing from the planet.
The disturbing function is
\begin{align}\label{eq:Rsec}
    \mathcal R &= \mu_{\rm p} n^2 a^2 \left[ \frac18\alpha b_{3/2}^{(1)}(\alpha) e^2 \right.\nonumber\\
    &\qquad\left.- \frac14\alpha b_{3/2}^{(2)}(\alpha)ee_{\rm p}\cos(\varpi_{\rm p}-\varpi)\right].
\end{align}
For simplicity, the Laplace coefficients are evaluated at $\alpha=\alpha_0$. 
For $j=2$, $b_{3/2}^{(1)}(\alpha_0)\approx 12.1$ and $b_{3/2}^{(2)}(\alpha_0)\approx 10.5$.
The secular effects are then calculated via Lagrange's planetary equations, $\dot \varpi_{\rm sec} = (na^2e)^{-1}\partial\mathcal R/\partial e$ and $\dot e_{\rm sec} = -(na^2 e)^{-1}\partial\mathcal R/\partial\varpi$.
Note that the apsidal angles $\varpi_{\rm p}$ and $\varpi$ satisfy $\varpi_{\rm p}-\varpi=\theta-\theta_{\rm p}$.

\subsection{Equations of Motion}
The Hamiltonian (\ref{eq:H}), along with the disturbing function (\ref{eq:Rsec}), yields the following equations of motion for the test particle,
\begin{align}
    &\dot n = ~3 \mu_{\rm p} \beta\alpha^{-1}  e n^{2} \left(j + 1\right) \sin{\theta} \nonumber\label{eq:ndot}\\
    &\qquad- 3\mu_{\rm p} \beta_{\rm p}  \alpha^{-1} e_{\rm p} n^{2} \left(j + 1\right) \sin{\theta_{\rm p}},\\
    &\dot e  = -  \mu_{\rm p}\alpha^{-1} \beta n \sin{\theta}+\frac12 \mu_{\rm p}\alpha^{-1} \beta e^{2}  n \left(j + 1\right) \sin{\theta} \nonumber\\
    &\qquad- \frac12 \mu_{\rm p}\alpha^{-1} \beta_{\rm p} e e_{\rm p} n \left(j + 1\right) \sin{\theta_{\rm p} }\nonumber\\
    &\qquad+\frac14 \mu_{\rm p}n\alpha b_{3/2}^{(2)}(\alpha_0)e_{\rm p}\sin(\omega_{\rm p}t-\varpi)\label{eq:edot},\\
    &\dot\varpi =~ \frac{\mu_{\rm p} \beta n }{\alpha e}\cos\theta + \omega+\frac14\mu_{\rm p} n \alpha b_{3/2}^{(1)}(\alpha_0) \nonumber\\
    &\qquad- \frac14\mu_{\rm p} n\alpha b_{3/2}^{(2)}(\alpha_0)\frac{e_{\rm p}}{e}\cos(\omega_{\rm p}t-\varpi)\label{eq:pomdot},\\
    &\dot\theta_{\rm p} = \left(j + 1\right) n - j n_{\rm p} \nonumber\\
    &\qquad+ \frac12  \mu_{\rm p} \alpha^{-1} \beta en \left(j + 1\right) \cos{\theta } - \omega_{\rm p}. \label{eq:thpdot}
\end{align}
Here we have used $n=\sqrt{GM/a^3}$ ($n$ is the osculating mean motion) and $\varpi_{\rm p}=\omega_{\rm p}t$.
The planet's mean motion $n_{\rm p}$ is decreased over time according to equation~(\ref{eq:npdot}).

In order to simulate resonance capture under the influence of differential apsidal precession, we integrate equations~(\ref{eq:ndot})--(\ref{eq:thpdot}) with the \texttt{scipy} Python package, using the Runge-Kutta method of order 8. 
We set the relative and absolute error tolerances to $10^{-9}$.

\subsection{Differential Apsidal Precession from Interior Planets}
For simplicity, we treat the externally induced apsidal precession rate $\omega$ and $\omega_{\rm p}$ as being constant, neglecting their variation in time.
In this paper, we consider the case where the external potential is generated by a quadrupole interior to the planet and the test particle.
This kind of quadrupole is induced by Jupiter, Saturn, and Uranus on Neptune and its resonant TNOs.
For Neptune, at its current location, $\omega_{\rm p}/n_{\rm p}=1.12\times10^{-4}$.
Using the approximation $b_{3/2}^{(1)}(\chi)\simeq 3\chi$ for $\chi\ll 1$ \citep{murraySolarSystemDynamics2000}, we find
\begin{equation}\label{eq:omomp}
\omega/\omega_{\rm p}\simeq (n/n_{\rm p})(a_{\rm p}/a)^2\simeq\alpha_0^{7/2}
\end{equation}
near the resonance.
Since $\alpha_0<1$, $\Delta\omega_{\rm ext} >0$.
We will use this relation in  Section~\ref{sec:capout}.

\subsection{Chaos Diagnostics}\label{sec:FLI}
We use the Fast Lyapunov Indicator \citep[FLI,][]{skokosChaosDetectionPredictability2016} to visualize chaotic zones in phase space.
When computing the FLI, we transform the time-dependent Hamiltonian (\ref{eq:H}) into an autonomous form using a time-dependent generating function to enable application of variational methods.
Then we integrate the variational equations of the Hamiltonian system of resonant dynamics using the same numerical integration scheme as for the equations of motion.
See Appendix~\ref{app:vareqs} for details.
As suggested in \citet{skokosChaosDetectionPredictability2016}, we utilize an orthonormal basis of tangent vectors and compute the average of the FLI in all directions. 

\begin{figure*}
    \centering
    \includegraphics[width=0.99\textwidth]{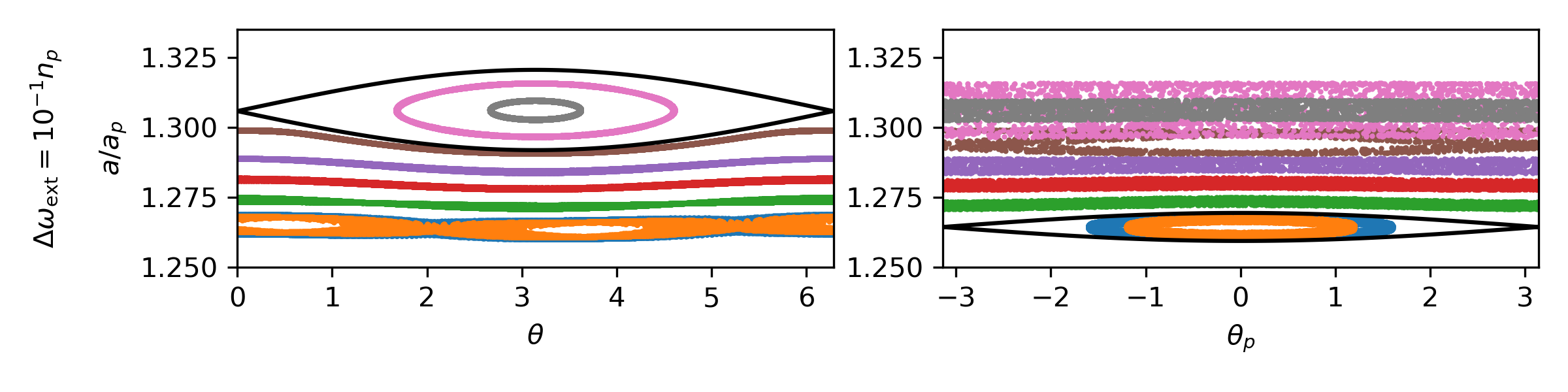}\\
    \includegraphics[width=0.99\textwidth]{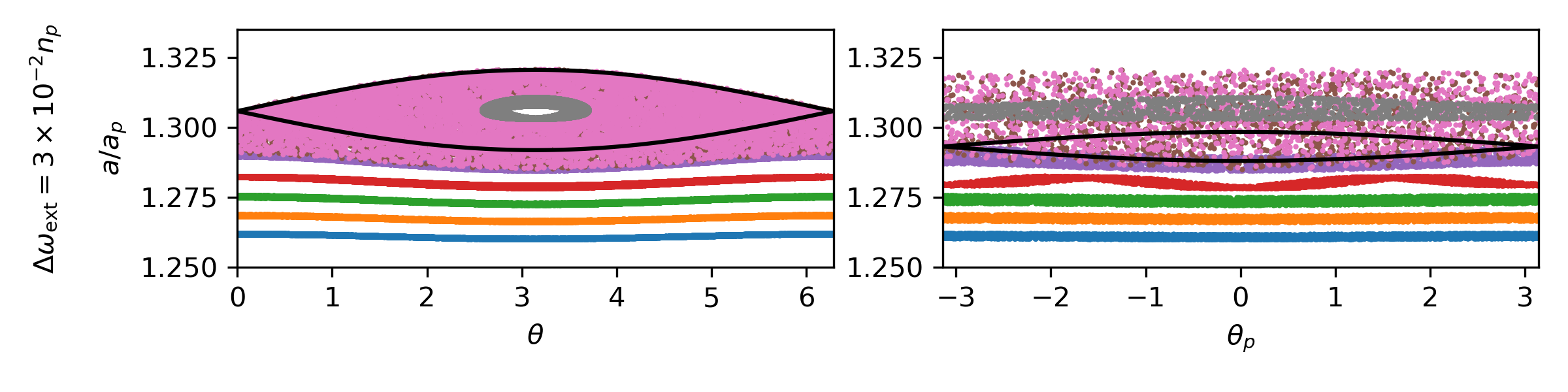}\\
    \includegraphics[width=0.99\textwidth]{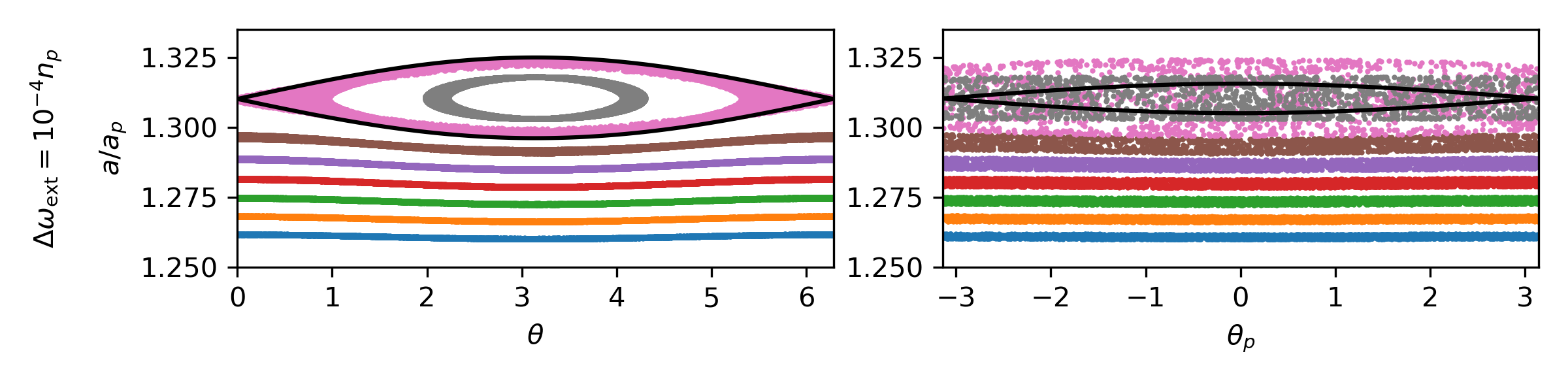}\\
    \includegraphics[width=0.99\textwidth]{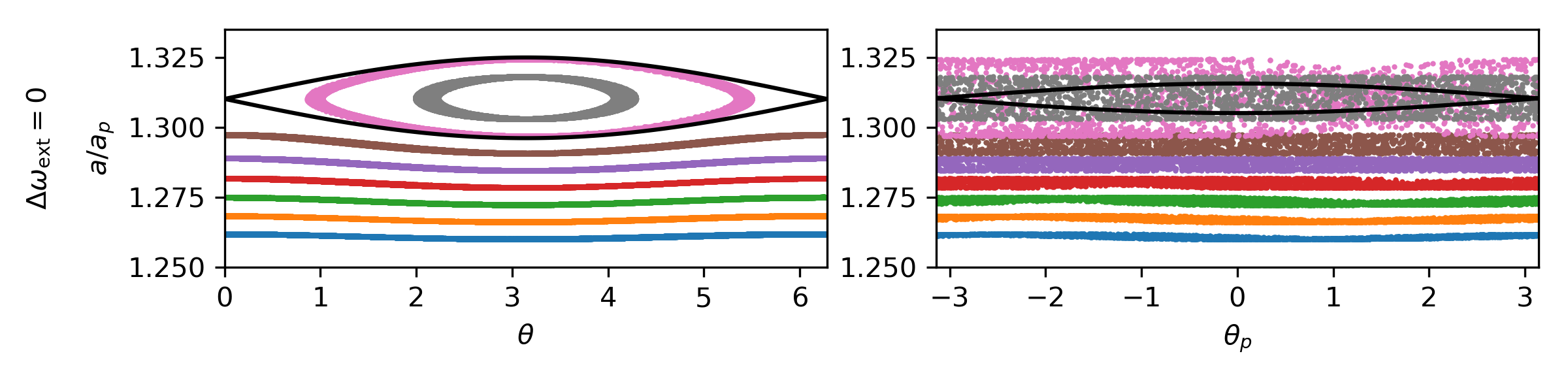}
    \caption{
    Phase-space trajectories of the 2:3 resonant system at varying degrees of differential apsidal precession, $\Delta\omega_{\rm ext}$, for $\mu_{\rm p}=5\times10^{-5}$ and $e_{\rm p}=0.03$. 
    The secular forcing from $m_{\rm p}$ on the test mass is neglected.
    Each color-coded trajectory corresponds to a different initial SMA.
    The left panels show $\theta$ vs. $a$; the right panels show $\theta_{\rm p}$ vs. $a$.
    The system is initialized with $k=-0.1$, which sets the initial $e$ via equation~(\ref{eq:k}).
    The initial separatrices for $\theta$ and $\theta_{\rm p}$ are plotted in black.
    The precession rate $\Delta\omega_{\rm ext}$ decreases from top to bottom: $0.1n_{\rm p}$ (row 1), $0.03n_{\rm p}$ (row 2), $10^{-4}n_{\rm p}$ (row 3), and 0 (row 4).
    }
    \label{fig:5e-5traj}
\end{figure*}
\begin{figure*}
    \centering
    \includegraphics[width=\textwidth]{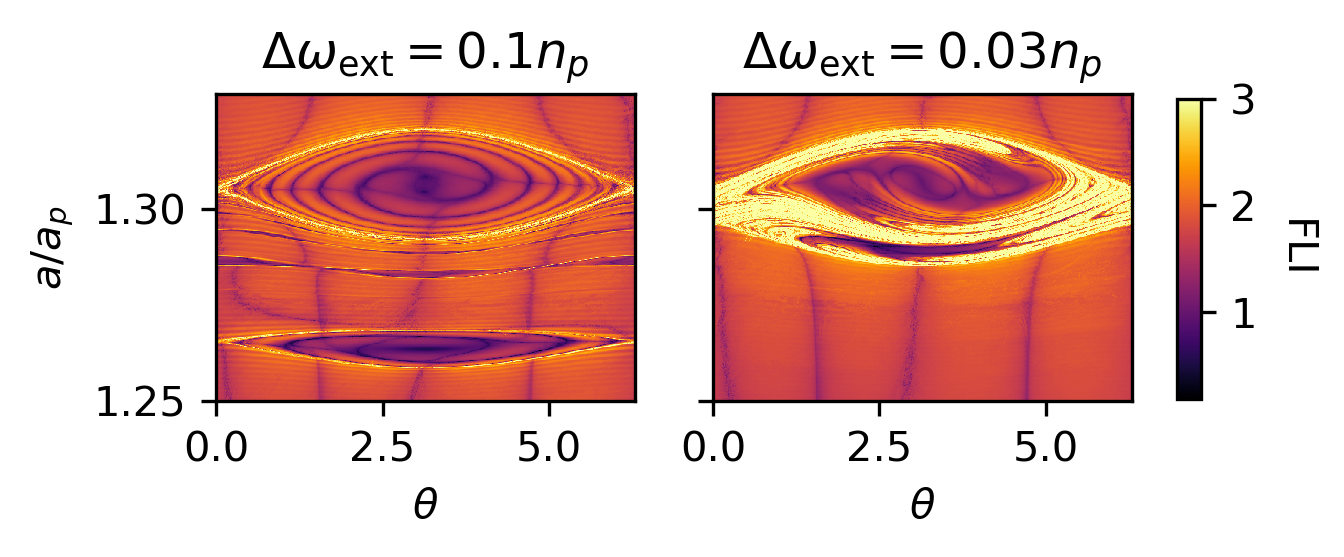}
    \caption{
    FLI (Section~\ref{sec:FLI}) as a function of the initial $a$ and $\theta$ for the resonant system (without the secular forcing from $m_{\rm p}$ and planet migration) for $\mu_{\rm p}=5\times10^{-5}$ and $e_{\rm p}=0.03$.
    The value of $k=-0.1$ is held fixed and $e$ is chosen via equation~(\ref{eq:k}).
    We initialize $\theta_{\rm p}=\theta+\pi$.
    \textit{Left:} $\Delta\omega_{\rm ext}=0.1n_{\rm p}$; the separatrices are clearly visible and well-separated.
    \textit{Right:} $\Delta\omega_{\rm ext}=0.03n_{\rm p}$; the separatrices overlap and widespread chaos permeates the resonant zone (high FLI values).
    }
    \label{fig:5e-5FLI}
\end{figure*}

\section{Resonant Dynamics and Resonance Overlap}\label{sec:resdyn}
To isolate the role that resonance overlap plays in the disruption of resonance capture, we first focus on a simplified system without migration ($n_{\rm p}$ is held constant). 
The external precession terms $\omega_{\rm p}$ and $\omega$ are treated as independent constants to precisely control the resonance centers.
We also neglect the secular action of the planet on the test particle, and so the resonant phase space trajectories are simple and clear.

\subsection{Isolated Resonances}
First, we consider the $\cos\theta$ resonance alone, i.e., we ignore the $\cos\theta_{\rm p}$ term in equation~(\ref{eq:H}).
Define the resonance center $n_c\equiv\sqrt{GM/a_c^3}$ via the relation
\begin{equation}\label{eq:nc}
    (j+1)n_{\rm c}-jn_{\rm p} -\omega = 0.
\end{equation}
If we scale the time by $n_{\rm p}^{-1}$ and the momenta by $\frac12\sqrt{GMa_{\rm c}}$, then for small $\delta a \equiv a-a_c$, the Hamiltonian for the test particle is 
\begin{align}\label{eq:calH}
    \mathcal H &= -\frac34 (j+1)(j+\omega/n_{\rm p}) e^4 - \frac32 k\left(j+\omega/n_{\rm p}\right)e^2 \nonumber\\
    &\quad- 2\beta\sqrt{a_{\rm p}/a_{\rm c}}\mu_{\rm p}e\cos\theta,
\end{align}
where
\begin{align}\label{eq:k}
    k\simeq \frac{\delta a}{a_{\rm c}} - (j+1) e^2
\end{align}
is a conserved quantity of the isolated resonance.
See Appendix~\ref{app:thH} for details.
The canonically conjugate momentum-coordinate pair is
\begin{equation}
    e^2\longleftrightarrow \theta.
\end{equation}
For $\omega=0$, equation~(\ref{eq:calH}) reduces to the standard Andoyer Hamiltonian, which is well-studied~\citep[e.g.][]{henrardSecondFundamentalModel1983a,murraySolarSystemDynamics2000}. 
The critical value of $k$ for which a separatrix appears, $k_{\rm sep}$, is
\begin{align}\label{eq:ksep}
    k_{\rm sep} = -\left(\frac{3(j+1)a_{\rm p}\beta^2}{j^2a_c}\right)^{1/3}\mu_{\rm p}^{2/3}.
\end{align}
We neglect $\omega$ from equation~(\ref{eq:ksep}) because $\omega/n_p\ll 1$.
For $k<k_{\rm sep}$, there is a separatrix that passes through the unstable fixed point at $\theta=0$.
Using the conservation of $\mathcal H$ and $k$, we can plot the separatrix on the $\theta-a$ phase plane for a given value of $k$ and $a$.
At $k_{\rm sep}$, the resonance width in SMA is (see Appendix~\ref{app:thH})
\begin{equation}\label{eq:dasep}
    \left|\frac{\Delta a_{\rm sep}}{a_c}\right| = \frac83 k_{\rm sep}.
\end{equation}
Note that it is straightforward to convert from SMA space to frequency space via the relation
\begin{align}
    \frac{\Delta n_{\rm c}}{n_c} \simeq \frac32 \frac{\Delta a_{\rm sep}}{a_c},
\end{align}
where $\Delta n_{\rm c}$ is the critical frequency width of the resonance when the separatrix first appears.
We estimate that resonance overlap first occurs at $\Delta\omega_{\rm ext}\simeq\Delta\omega_{\rm crit}$, where
\begin{align}\label{eq:Domcrit}
\Delta\omega_{\rm crit}\equiv \Delta n_{\rm c}.
\end{align}

Now instead consider the $\cos\theta_{\rm p}$ term in equation~(\ref{eq:H}) alone. 
Define the resonant center at $n_{c,p}\equiv(GM/a_{c,p}^3)^{1/2}$ via the relation
\begin{equation}\label{eq:nc}
    (j+1)n_{\rm c,p}-jn_{\rm p} -\omega_{\rm p} = 0.
\end{equation}
The system is a simple pendulum,
\begin{align}
    \ddot\theta_{\rm p} = -\frac{3(j+1)^2\beta_{\rm p}}{\alpha}n_{\rm c,p}^2\mu_{\rm p}e_{\rm p}\sin\theta_{\rm p}.
\end{align}
The half-width of the separatrix in SMA space is 
\begin{align}\label{eq:dasepp}
    \frac{\Delta a_{\rm sep,p}}{a_{c,p}}\approx 4\sqrt{\frac{\beta_{\rm p}\mu_{\rm p} e_{\rm p}a_{\rm c,p}}{3a_{\rm p}}},
\end{align}
where $\Delta a_{\rm sep,p} \equiv a-a_{c,p}$ is assumed to be small.
In the frequency space, the width is
\begin{align}
    \frac{\Delta n_{\rm c,p}}{n_{\rm c,p}} \approx \frac32\frac{\Delta a_{\rm sep,p}}{a_{c,p}}.
\end{align}

\subsection{Phase-space Trajectories}
We examine system trajectories for various values of $\Delta\omega_{\rm ext}$ for $\mu_{\rm p}=5\times10^{-5}$ and $e_{\rm p}=0.03$ in Figure~\ref{fig:5e-5traj}.
We choose 8 initial values of $a$ uniformly sampled between 1.26 and 1.303 and initialize $\theta_{\rm p}=0$ and $\varpi-\varpi_{\rm p}=\pi$.
The runs with the same initial conditions are color-coded consistently between the left and right panels. 
We set $k=-0.1$ at $t=0$, which fixes the initial shape of the $\theta$ separatrix, depicted as a black line (left panels).
The initial $e$ is determined by the choices for $k$ (equation~\ref{eq:k}) and $a$.
This separatrix is only formally well-defined if $\theta$ is considered in isolation.
Here, we use it to approximate the region of influence of $\theta$ and to illustrate resonance overlap.
Likewise, the separatrix for $\theta_{\rm p}$ (at $t=0$) is drawn on the phase portrait in the right panels.
We integrate the system forward in time, and record the dynamical variables at every perihelion passage of the planet ($t=0,2\pi,4\pi,\ldots)$.

At the top of Figure~\ref{fig:5e-5traj}, we consider a very large value of $\Delta\omega_{\rm ext}=0.1n_{\rm p}$.
In this case, the separatrices for $\theta$ and $\theta_{\rm p}$ are well-separated.
There are trajectories trapped in the $\theta$ resonance, indicated by librating $\theta$ angles (pink and gray).
There are also trajectories trapped in the $\theta_{\rm p}$ resonance in the right panel (orange and blue).
In the second row of Figure~\ref{fig:5e-5traj}, the value of $\Delta\omega_{\rm ext}$ is chosen so that the two separatrices cross.
The separation of the resonance centers is comparable to both of the separatrices' widths, and resonance overlap occurs.
In this case, trajectories that were formerly trapped within and near the $\theta$ resonance (pink and brown) circulate in $\theta$, and their motion appears irregular.
The trajectories that are central to the $\theta$ resonance remain trapped (gray).
In the third row, we choose a small value of $\Delta\omega_{\rm ext}=10^{-4}n_{\rm p}$.
We find that this value of $\Delta\omega_{\rm ext}$, despite being much smaller than the resonance widths, is still enough to destabilize formerly resonant trajectories:
The outermost trajectory (pink) in the $\theta$ separatrix for $\Delta\omega_{\rm ext}=0$ (bottom row) is ``stretched'' out to the corners of the separatrix and becomes circulating.
We hypothesize that this mechanism contributes to capture disruption, as discussed further in Section~\ref{sec:resoverlap}.
In the bottom row, we set $\Delta\omega_{\rm ext}=0$.
The resonances are superimposed and the problem is integrable (equation~\ref{eq:Hbar}).
In this case, $e_{\rm p}$ is small enough so that the $\theta$ resonance still librates~\citep{launeApsidalAlignmentAntialignment2022}, while $\theta_{\rm p}$ circulates.

\subsection{FLI Mapping}
In Figure~\ref{fig:5e-5FLI}, we map out the resonance in phase space with the FLI (see Section~\ref{sec:FLI} and Appendix~\ref{app:vareqs}).
We set up runs in a 400$\times$400 grid of the initial $a/a_{\rm p}$ and $\theta$.
We again select $k=-0.1$ and $e_{\rm p}=0.03$ and  initialize $\varpi-\varpi_{\rm p}=\pi$, which sets the alignment of the $\theta_{\rm p}$ separatrix on the $\theta$ axis.
In the left panel, we choose $\Delta\omega_{\rm ext}=0.1n_{\rm p}$, so that the $\theta$ and $\theta_{\rm p}$ separatrices are well-separated.
In this case, the two separatrices appear clearly near the predicted locations (c.f. the top panel of Figure~\ref{fig:5e-5traj}).
The $\theta_{\rm p}$ separatrix appears to be slightly distorted from the isolated case due to interaction with the $\theta$ resonance.
The right panel of Figure~\ref{fig:5e-5FLI} has $\Delta\omega_{\rm ext}=0.03n_{\rm p}$, so that the separatrices cross near $\theta=\pi$ (c.f. the second panel of Figure~\ref{fig:5e-5traj}).
Widespread chaos, indicated by elevated FLI values, permeates the resonant zone.
There are islands of low FLI in the central regions of the original separatrices.
These features corroborate our findings in Figure~\ref{fig:5e-5traj}.

\subsection{Combined Resonances}\label{sec:combres}
In the absence of the differential apsidal precession ($\Delta\omega_{\rm ext}=0$), the angles $\theta$ and $\theta_{\rm p}$ can be combined into a single mixed resonance angle via a series of canonical transformations, thereby reducing the system to a single degree of freedom and rendering it integrable~\citep[c.f.][]{wisdomCanonicalSolutionTwo1986,henrardReducingTransformationApocentric1986,deck13_first_order_reson_overl_stabil,elmoutamidCouplingCorotationLindblad2014,petitResonanceK219System2020,launeApsidalAlignmentAntialignment2022}.
The elevated FLI in Figure~\ref{fig:5e-5FLI}, caused by resonance overlap, indicates that reducing the system to a single degree of freedom should fail whenever $\Delta\omega_{\rm ext} \neq 0$.
We now examine how this reduction fails.

Assuming $\alpha\simeq\alpha_0$, we perform an analogous series of canonical transformations, this time accounting for the apsidal precession terms.
We provide details in Appendix~\ref{app:canontrans}.
The resulting Hamiltonian, valid near resonance, is
\begin{align}\label{eq:Hbar}
    \overline{\mathcal{H}} =& 
    -\frac32\left(\frac{j+\omega_{\rm p}/n_{\rm p}}{j+1}\right)^{4/3}\delta\Theta^2+\Delta\omega_{\rm ext}\Phi\nonumber\\
    &-\alpha_0^{1/4} \beta\mu_{\rm p}\sqrt{2\Phi}\cos(\theta_{\rm p}+\phi)\nonumber\\
    &+\frac{\beta_{\rm p}}{\beta\alpha_0^{1/4}}e_{\rm p}\Delta\omega_{\rm ext}\sqrt{2\Phi}\cos\phi.
\end{align}
The action-angle variables are
\begin{align}\label{eq:Theta}
    \delta\Theta&=\sqrt{\frac{a}{a_{\rm p}}}-\sqrt{\frac{a_{c,p}}{a_{\rm p}}}\nonumber\\
    &\longleftrightarrow \frac{\theta_{\rm p}}{j+1}=\lambda-\left(\frac{j+\omega_{\rm p}/n_{\rm p}}{j+1}\right)t,
\end{align}
\begin{align}
    \Phi
    &=\frac{1}{2\sqrt{\alpha_0}}\norm{\mathbf e - \left(\beta_{\rm p}/\beta\right)\mathbf e_{\rm p}}^2\nonumber\\
    &=\frac{e^2}{2\sqrt{\alpha_0}}-\frac{\beta_{\rm p}}{\beta\sqrt{\alpha_0}}ee_{\rm p}\cos(\omega_{\rm p}t-\varpi)+\frac{e_{\rm p}^2\beta_{\rm p}^2}{2\sqrt{\alpha_0}\beta^2}\nonumber\\
    &\longleftrightarrow \phi=\arctan\left[\frac{e\sin(\omega_{\rm p}t-\varpi)}{e\cos(\omega_{\rm p}t-\varpi)-(\beta_{\rm p}/\beta)e_{\rm p}}\right],\label{eq:phi}
\end{align}
where $\mathbf e$ and $\mathbf e_{\rm p}$ are the Runge-Lenz vectors of the test mass and planet, respectively.
Generally we work directly with the orbital elements and the Hamiltonian in the form of equation~(\ref{eq:H}). 
But equation~(\ref{eq:Hbar}) is written in action-angle coordinates adapted to the differential apsidal precession.
One can see that as $\Delta\omega_{\rm ext}\to 0$, the problem reduces to the familiar single resonance situation and is therefore integrable.

\begin{figure*}
    \centering
    \includegraphics[width=0.48\textwidth]{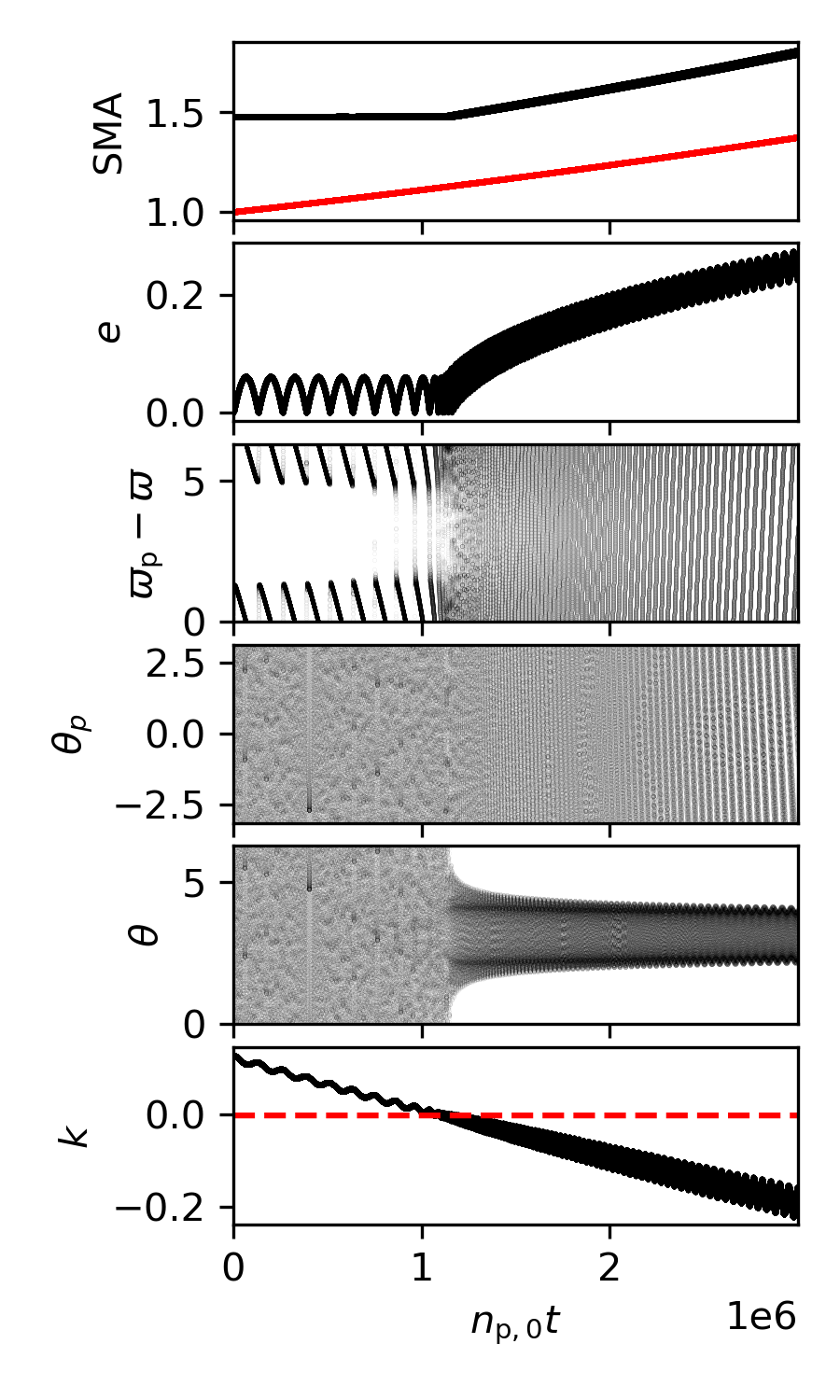}
    \includegraphics[width=0.48\textwidth]{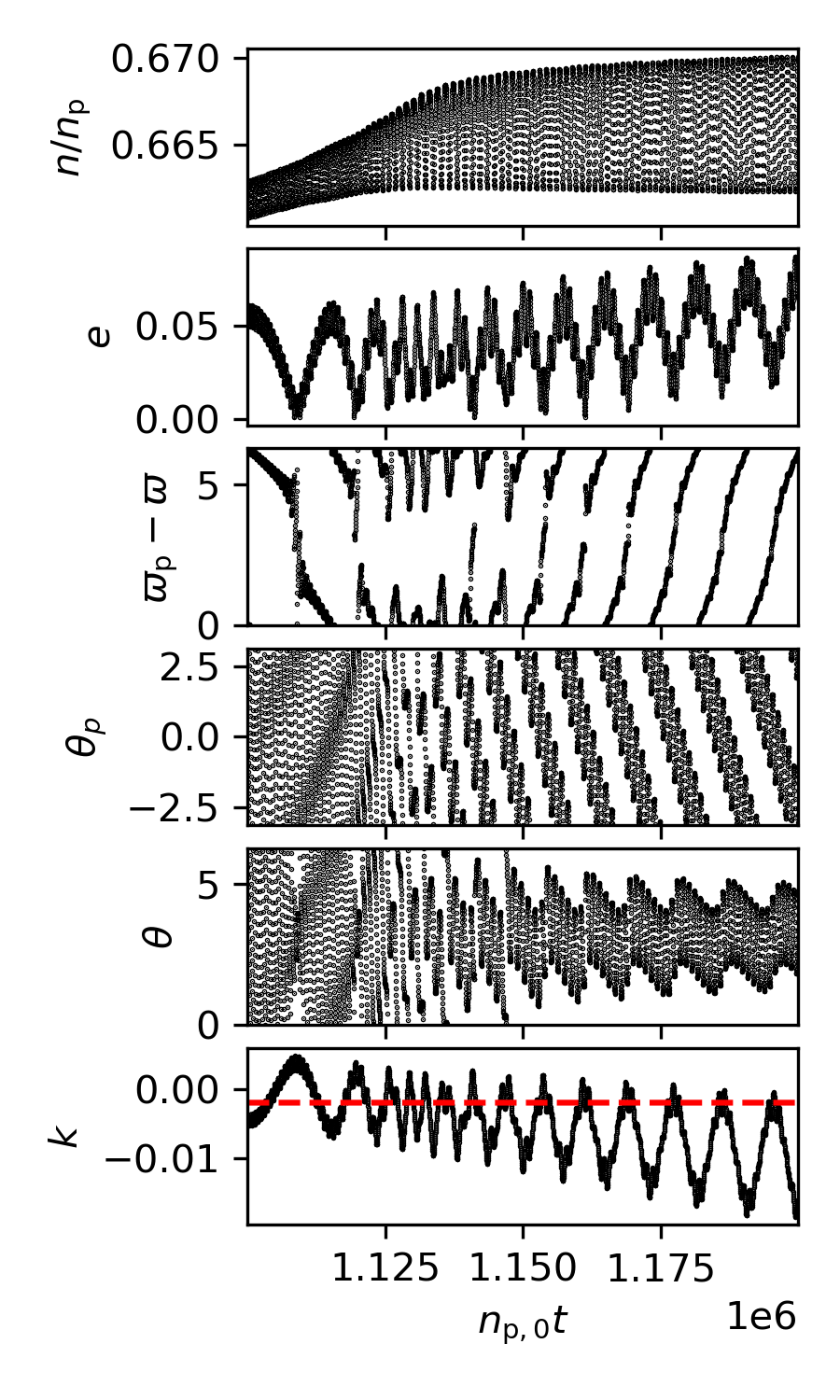}
    \caption{Capture of the test particle into a 2:3 MMR with a planet of mass $\mu_{\rm p}=5\times10^{-5}$ and $e_{\rm p}=0.03$, under a weak differential apsidal precession ($\Delta\omega_{\rm ext}=10^{-5}n_{\rm p,0}$).
    The initial conditions of the test particle are $\theta_{\rm p}=0$, $e=0.001$, $\varpi=\pi$ and $a/a_{\rm p}\simeq1.48$ ($n/n_{\rm p,0}=1/1.8$).
    As the planet migrates outward, $n/n_{\rm p}$ increases and approaches $2/3$.
    When $k$ (equation~\ref{eq:k}) crosses $k_{\rm sep}$ (equation~\ref{eq:ksep}, red line), $\theta$ begins to librate and $e$ grows as the system is transported deeper into resonance.
    \textit{Left:} Full time evolution. In the top panel, the red line indicates $a_{\rm p}$.
    \textit{Right:} Zoomed-in view near the moment of $\theta$ capture.
    }
    \label{fig:capex}
\end{figure*}
\begin{figure*}
    \centering
    \includegraphics[width=0.48\textwidth]{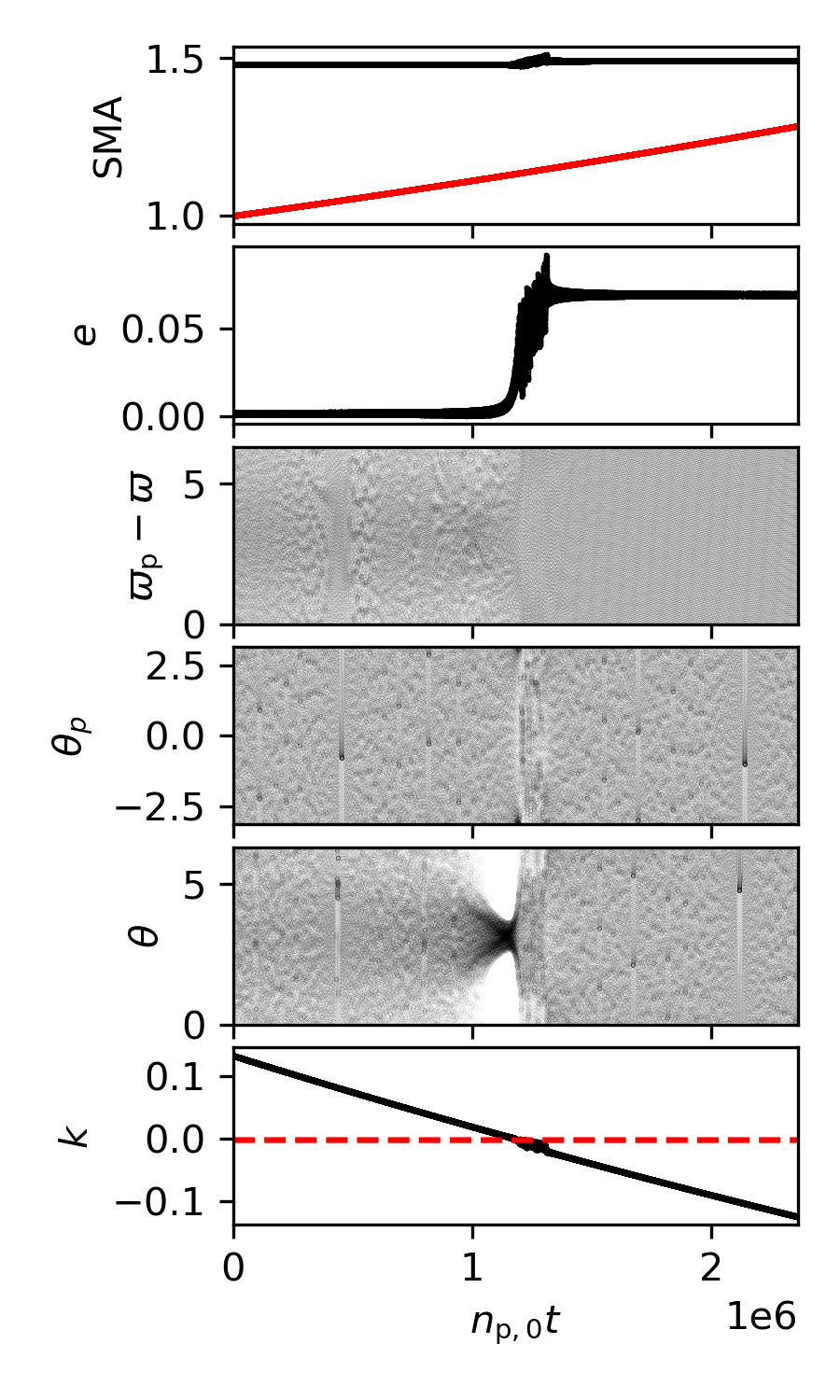}
    \includegraphics[width=0.48\textwidth]{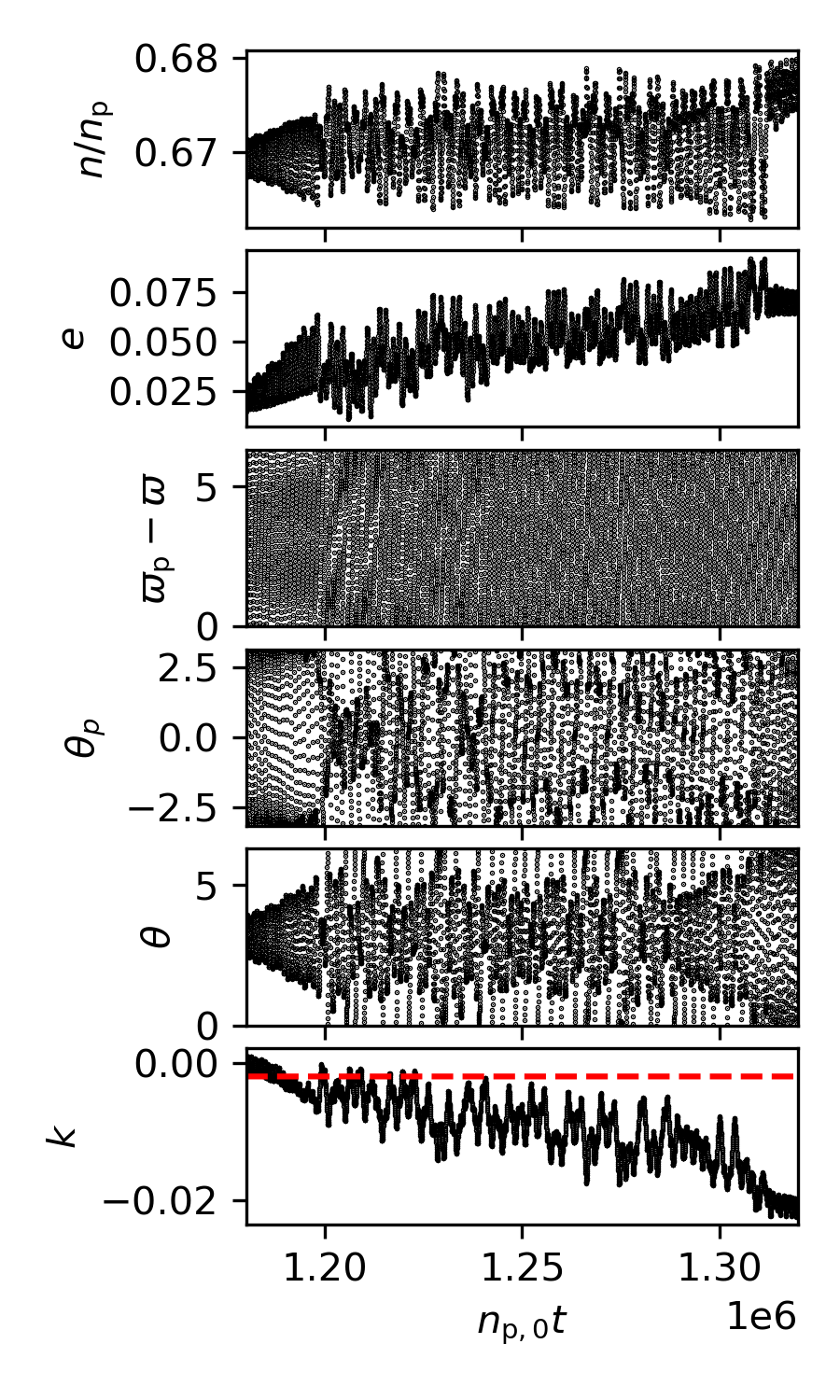}
    \caption{
    Same as Figure~\ref{fig:capex}, but with $\Delta\omega_{\rm ext}=5\times 10^{-3}n_{\rm p,0}$. 
    The test particle passes through the resonance, but its eccentricity remains low.
    Both angles $\theta$ and $\theta_{\rm p}$ alternate chaotically between libration and circulation until the particle exits the resonance.
    \textit{Left:} Full time evolution.
    \textit{Right:} Zoomed-in view near the moment of disruption.
    }
    \label{fig:disruptoverlapex}
\end{figure*}
\begin{figure*}
    \centering
    \includegraphics[width=0.48\textwidth]{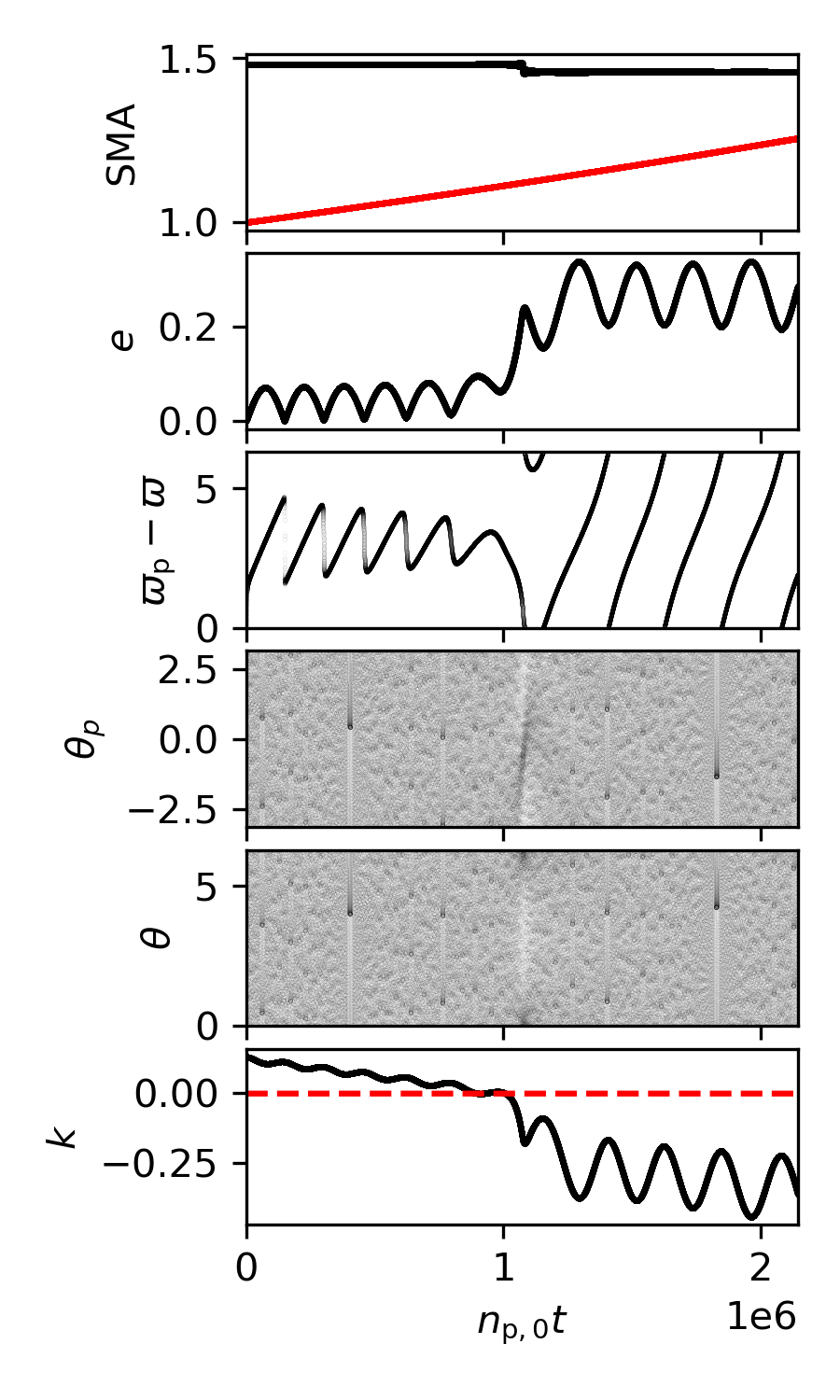}
    \includegraphics[width=0.48\textwidth]{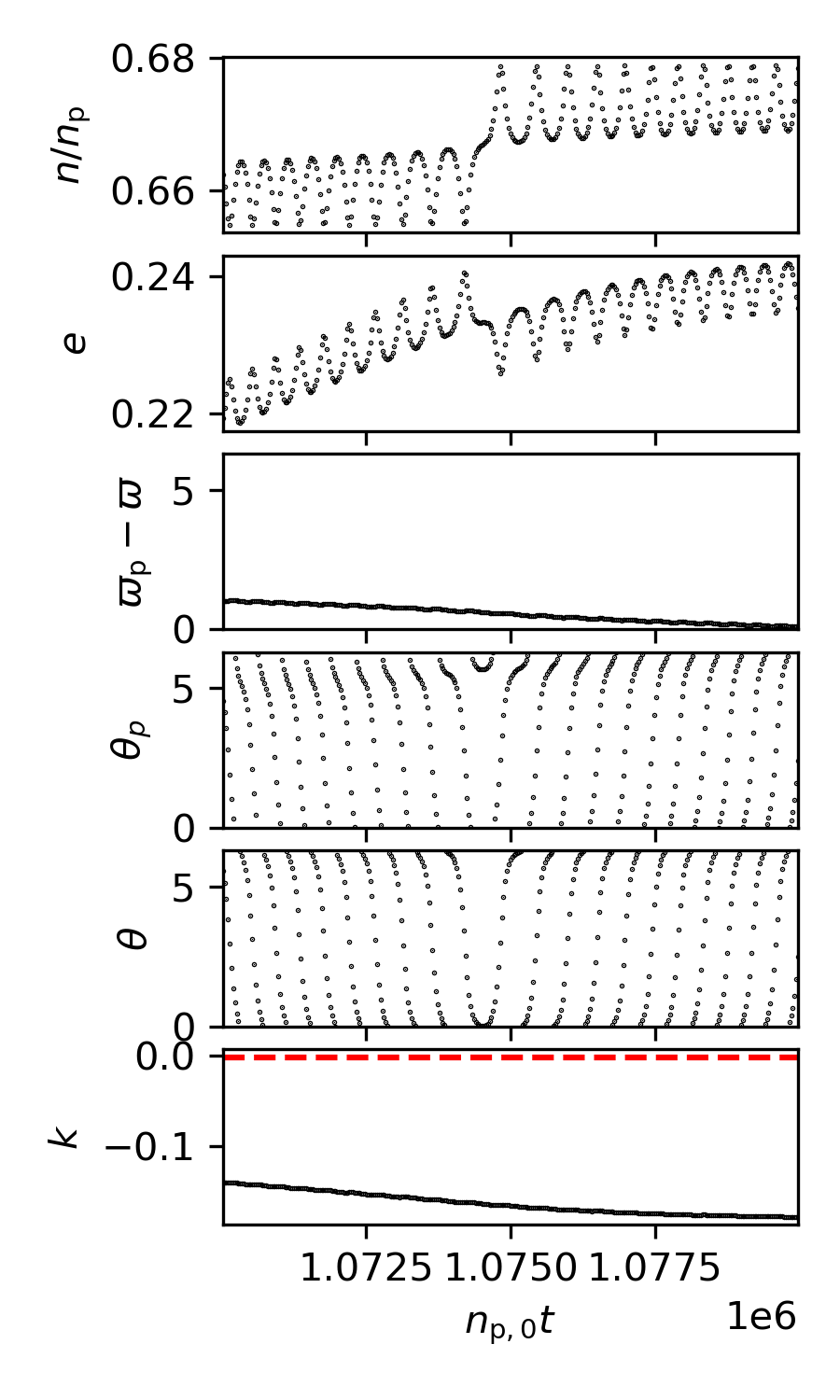}
    \caption{
    Same as Figure~\ref{fig:capex}, but for $\Delta\omega_{\rm ext}=10^{-4}n_{\rm p,0}$. 
    The test particle passes through the resonance without capture; $\theta$ and $\theta_{\rm p}$ never enter libration.
    Around the resonance crossing, the particle's eccentricity increases to become orbit-crossing with the planet, making the system dynamically unstable and likely leading to ejection.
    \textit{Left:} Full time evolution.
    \textit{Right:} Zoomed-in view near the moment of disruption.
    }
    \label{fig:disruptex}
\end{figure*}
\begin{figure*}
    \centering
    \includegraphics[width=0.48\textwidth]{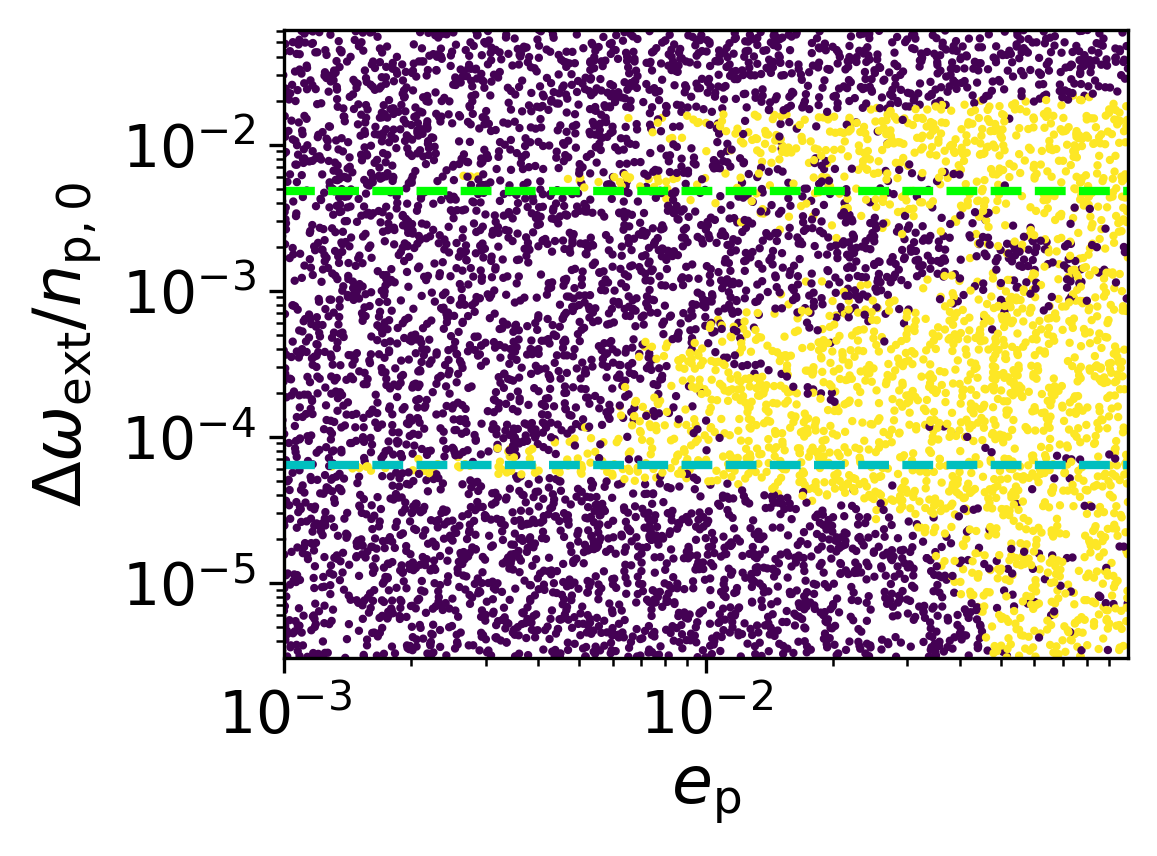}
    \includegraphics[width=0.48\textwidth]{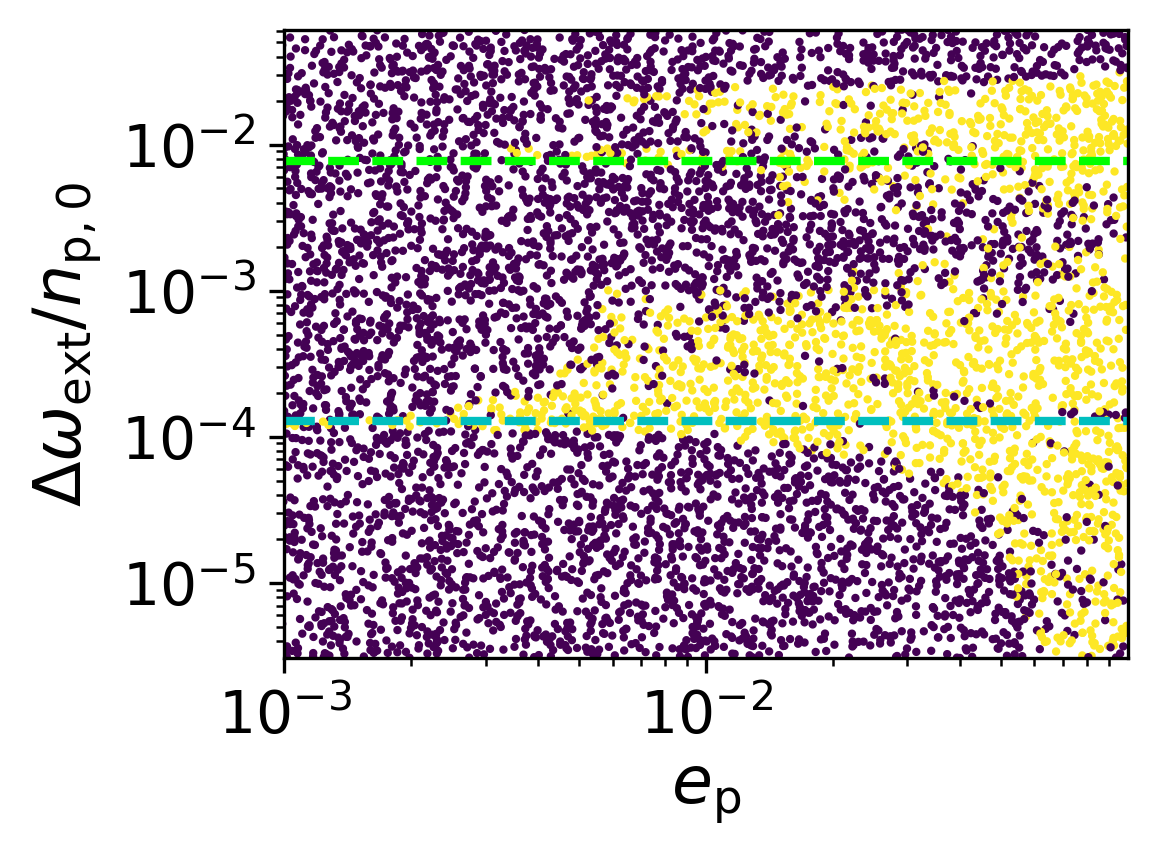}
    \caption{
    Capture outcomes for the $2:3$ MMR as a function of differential apsidal precession rate $\Delta\omega_{\rm ext}$ and planet eccentricity $e_{\rm p}$.
    Yellow dots indicate disruption; purple dots indicate successful capture. 
    The  resonance overlap threshold $\Delta \omega_{\rm crit}$ (equation~\ref{eq:Domcrit}) and secular resonance frequency $\Delta\omega_{\rm crit,sec}$ (equation~\ref{eq:Domcritsec}) are marked with green and blue dashed lines, respectively.
    \textit{Left:} $\mu_{\rm p}=5\times 10^{-5}$.
    \textit{Right:} $\mu_{\rm p}=10^{-4}$.
    }
    \label{fig:capgrid}
\end{figure*}

\section{Migration and Secular Effects}\label{sec:Migration}
Now we simulate the capture outcome for the test particle as the planet's $j:j+1$ resonance sweeps through its SMA by integrating equations~(\ref{eq:ndot}--\ref{eq:thpdot}).
We include the migration of $m_{\rm p}$ and the secular forcing of the planet on the test particle.

In Figure~\ref{fig:capex}, we show an example of successful capture into the $2:3$ $\theta$-resonance for weak apsidal precession, $\Delta\omega_{\rm ext}=10^{-5}n_{\rm p,0}$, with $\mu_{\rm p}=5\times10^{-5}$ and $e_{\rm p}=0.03$.
The test particle's SMA $a$ is initialized outside the resonance value $a_c$.
As the planet's orbit expands (equation~\ref{eq:npdot}), $a_c$ moves outward over time.
Before encountering the resonance, the eccentricity $e$ and apsidal angle $\varpi_{\rm p}-\varpi$ undergo secular oscillations, with $\varpi_{\rm p}-\varpi$ centered around $0$.
The quantity $k$ decreases over time due to the planet's migration; its short-term oscillations are due to the coupling with $e_{\rm p}$. 
As $k$ crosses the value $k_{\rm sep}$, the angle $\theta$ begins to librate and the particle enters into the resonance.
As the particle is adiabatically transported deeper into resonance, $e$ grows, $k$ decreases, and $a$ is trapped close to $a_c$.
When $\theta$ begins to librate, the apsidal angle $\varpi_{\rm p}-\varpi$ begins to circulate \citep{launeApsidalAlignmentAntialignment2022}.

In Figure~\ref{fig:disruptoverlapex}, we show the same integration as in Figure~\ref{fig:capex}, but for $\Delta\omega_{\rm ext}=5\times10^{-3}n_{\rm p,0}$.
Here we see the $\theta$ angle begin to librate as $k$ crosses the critical $k_{\rm sep}$ value.
Then, $\theta$ and $\theta_{\rm p}$ both alternate between circulation and libration, and the evolution of $n$ and $e$ appears chaotic.
After around $20,000$ orbits, the particle exits the resonance.
The eccentricity is relatively small, $e\sim 0.05$, and the apsidal angle $\varpi_{\rm p}-\varpi$ is circulating the entire time.

In Figure~\ref{fig:disruptex}, we show the same integration as in Figure~\ref{fig:capex}, but for $\Delta\omega_{\rm ext}=10^{-4}n_{\rm p,0}$.
Before encountering the resonance, the secular behavior of the system is largely the same, except that the apsidal angle $\varpi_{\rm p}-\varpi$ oscillates about $\pi$.
As the mean motion ratio $n/n_{\rm p}$ approaches the resonant value $n_c/n_{\rm p}\simeq2/3$, the eccentricity is excited to a large value, $e\approx0.23$, and $k$ is also driven to a larger magnitude, $k\simeq-0.16$.
When the test particle encounters the resonance, the angles $\theta$ and $\theta_{\rm p}$ simultaneously reverse course and $n$ ``skips'' through the resonance value $n_c$.
Capture does not occur and $n/n_{\rm p}$ continues to increase.
Near the resonance, when $e$ is excited, the perihelion distance of the test particle is smaller than the aphelion of the planet, $a(1-e)<a_{\rm p}(1+e_{\rm p})$.
This configuration is likely unstable and, in a real system, could lead to a close encounter and ejection of the test particle.

\subsection{Capture outcomes}\label{sec:capout}
In Figure~\ref{fig:capgrid}, we show the outcomes of convergent migration for a range of system parameters $e_{\rm p}$ and $\Delta\omega_{\rm ext}$ for the planet masses $\mu_{\rm p}=5\times10^{-5}$ (left) and $10^{-4}$ (right).
The planet precession rates $\omega_{\rm p}$ are chosen log-uniformly between $5\times10^{-6}n_{\rm p,0}$ and $0.1n_{\rm p,0}$, and $\omega$ and $\Delta\omega_{\rm ext}$ are set by equation~(\ref{eq:omomp}).
The planet eccentricity $e_{\rm p}$ is chosen log-uniformly between $10^{-3}$ and $0.1$.
The test particle is initialized with $e=0.001$, $n/n_{\rm p,0}=1/1.8$, and $\varpi$ chosen uniformly at random.
The planet has initial $\varpi_{\rm p}=0$, and the initial $\theta_{\rm p}$ is chosen uniformly at random.
The migration timescale is $\tau_m=10^6P_{\rm p, 0}$.

We see in Figure~\ref{fig:capgrid} that capture is secure for very small $e_{\rm p}\text{'s}$, regardless of the $\Delta\omega_{\rm ext}$  values.
For larger $e_{\rm p}\text{'s}$, even when $\Delta\omega_{\rm ext}$ is zero (not shown) or very small, capture is disrupted; 
this disruption occurs for $e_{\rm p}$ greater than some critical value, around $e_{\rm p}\approx 0.04$ and $e_{\rm p}\approx0.06$ for $\mu_{\rm p}=5\times10^{-5}$ and $\mu_{\rm p}=10^{-4}$, respectively.
For very large differential precession, capture is secure because the two separatrices are well separated. 
In between these two extremes, we see two distinct ``lobes'' in which capture is disrupted at small $e_{\rm p}$.
One lobe is in the vicinity of $\Delta\omega_{\rm ext}\sim 0.01n_{\rm p,0}$ and the other in the vicinity of $\Delta\omega_{\rm ext}\sim 10^{-4}n_{\rm p,0}$.
We argue that these features are due to two distinct processes: resonance overlap at stronger precession (see equation~\ref{eq:Domcrit} and Figure~\ref{fig:disruptoverlapex}) and a secular resonance at weaker precession (see equation~\ref{eq:Domcritsec} below and Figure~\ref{fig:disruptex}).

\subsection{Resonance overlap}\label{sec:resoverlap}
In Figure~\ref{fig:capgrid} we have indicated the (frequency) width $\Delta\omega_{\rm crit}$ of the $\theta$-separatrix when it first appears (equation~\ref{eq:Domcrit}) with a  green line.
We see that this line aligns well with the bottom of the upper lobe and with the ``spike'' of disruption there.
The integration for $\Delta\omega_{\rm ext}=5\times10^{-3}n_{\rm p,0}$ in Figure~\ref{fig:disruptoverlapex} also supports that these disruptions are due to resonance overlap.
The apsidal angle $\varpi_{\rm p}-\varpi$ (present in the secular equations) is always circulating, and $\theta$ and $\theta_{\rm p}$ exhibit coupled chaotic evolution until the test particle exits resonance.

\begin{figure}
    \centering
    \includegraphics[width=0.38\textwidth]{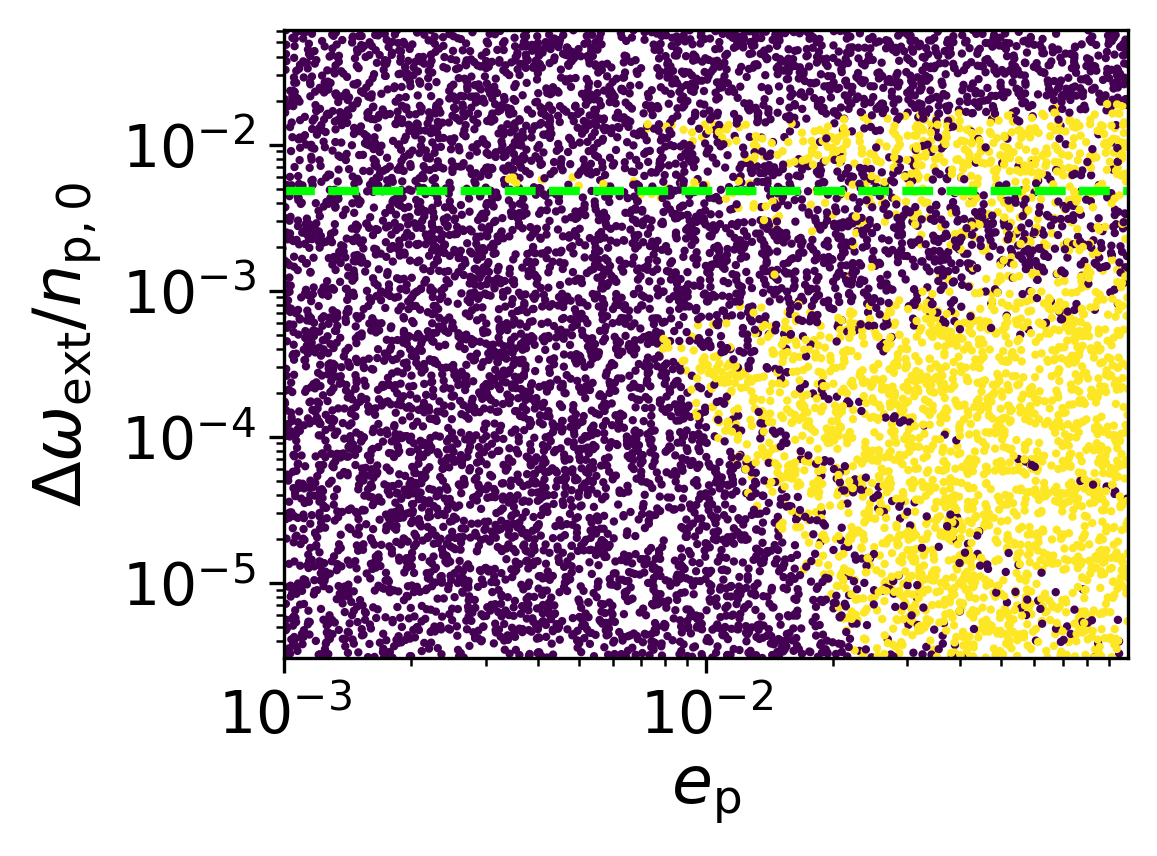}\\   \includegraphics[width=0.38\textwidth]{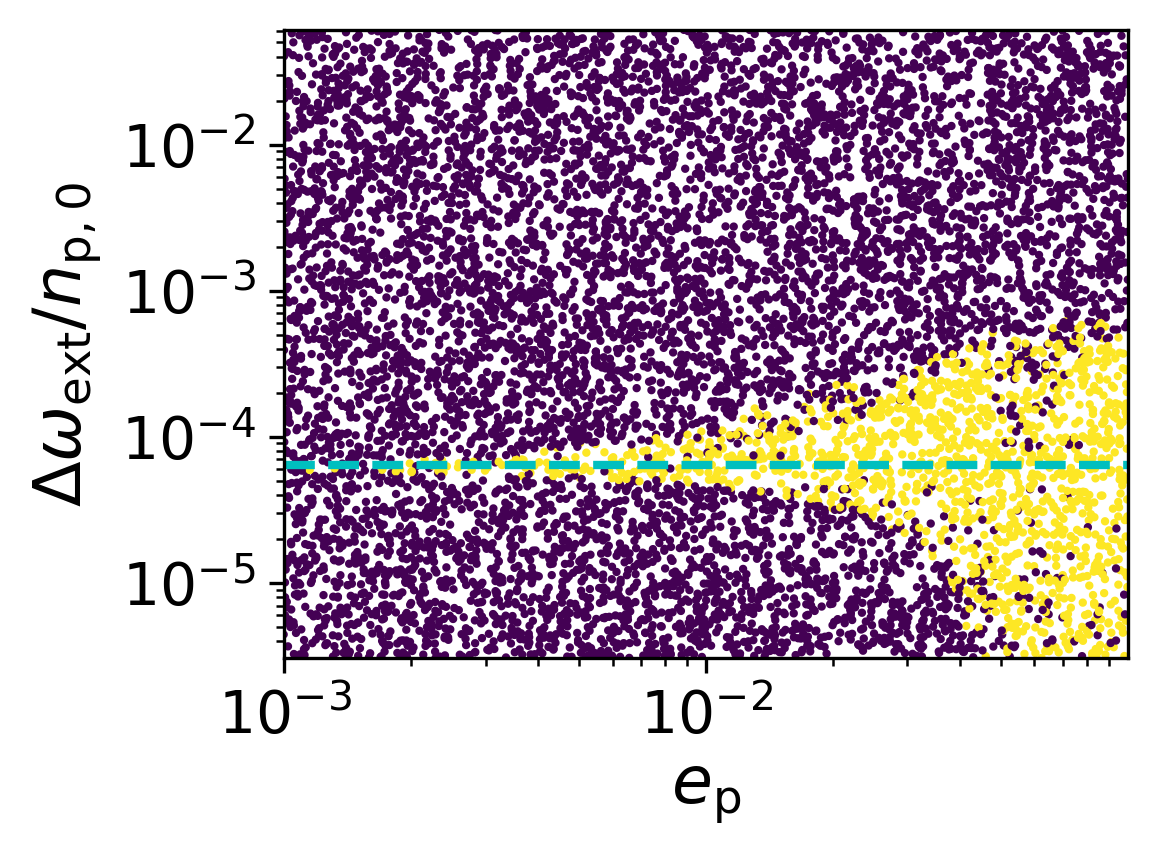}\\
    \includegraphics[width=0.38\textwidth]{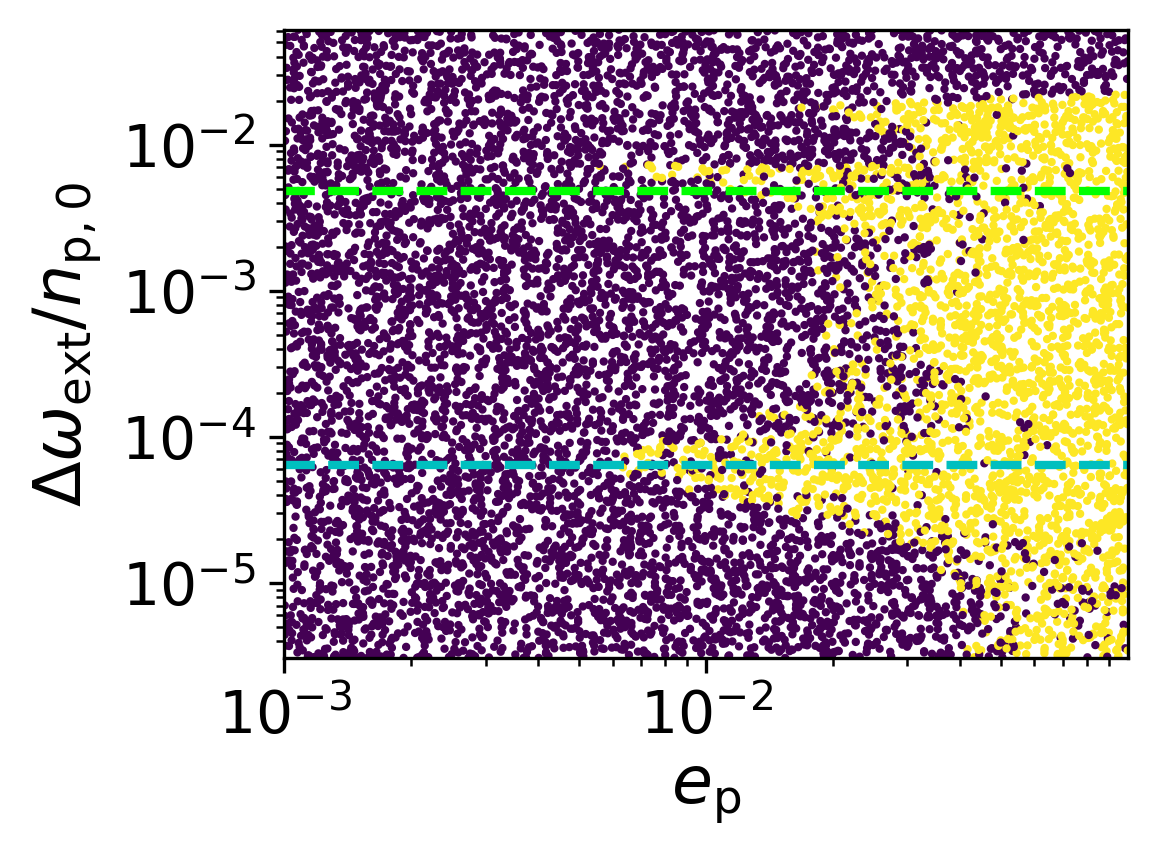}
    \caption{
    Same as Figure~\ref{fig:capgrid}, but for three diagnostic scenarios, all with $\mu_{\rm p}=5\times10^{-5}$. 
    \textit{Top:} Secular forcing from the planet on the test particle is neglected.
    \textit{Middle:} Resonant interaction involving $\theta_{\rm p}$ is disabled ($\beta_{\rm p}$ is set to $0$).
    \textit{Bottom:} Migration is ten times faster, with $\tau_m=10^{5}P_{\rm p,0}$.
    }
    \label{fig:ressectm}
\end{figure}

In the top panel of Figure~\ref{fig:ressectm}, we examine the capture problem for $\mu_{\rm p}=5\times10^{-5}$ but neglect the secular forcing from the planet on the test particle.
The top lobe is indeed still present, including the spike around $\Delta\omega_{\rm crit}$.
At smaller $\Delta\omega_{\rm ext}$, disruptions occur when $e_{\rm p}$ is large enough, but these do not depend strongly on the precession rate and there is no spike around $\Delta\omega_{\rm ext}\sim 6\times 10^{-5} n_{\rm p,0}$ as in Figure~\ref{fig:capgrid}.
We hypothesize that this feature is due to the dynamics presented in Figure~\ref{fig:5e-5traj}, where even for the very weak precession of $\Delta\omega_{\rm ext}=10^{-4}n_{\rm p}$ (see the third row of Figure~\ref{fig:5e-5traj}), the outer trajectories inside the $\theta$ separatrix are destabilized, and $\theta$ begins to circulate.

\subsection{Secular Resonance}\label{sec:SecRes}
In the middle panel of Figure~\ref{fig:ressectm}, we examine the capture problem for $\mu_{\rm p}=5\times10^{-5}$ but artificially set $\beta_{\rm p}=0$ in equations (\ref{eq:ndot}--\ref{eq:thpdot}) to isolate the effects of the secular coupling with $\mu_{\rm p}$.
Here we see the spike in disruption that is absent in the resonant-only runs (top panel).
If we compare Figures~\ref{fig:capex} (capture)  and \ref{fig:disruptex} (disruption), we notice that the apsidal angle $\varpi_{\rm p}-\varpi=\theta-\theta_{\rm p}$ librates around $0$ and $\pi$, respectively.
Another key difference is that the eccentricity is excited to larger values in the case of disruption.

Far from the resonance, both $\theta$ and $\theta_{\rm p}$ circulate, causing their resonant terms to average out to zero.
As a result, $\dot n=0$ (equation~\ref{eq:ndot}) and $\dot e=0$ (equation~\ref{eq:edot}).
The critical equation for equilibrium far from the resonance is therefore the secular terms in equation~(\ref{eq:pomdot}). 
Setting $\dot\varpi_{\rm p}-\dot\varpi=0$, and solving for the equilibrium eccentricity value, we find
\begin{align}\label{eq:eeq}
    e_{\rm eq,sec} = \frac{
    \frac14\alpha b_{3/2}^{(2)}(\alpha_0) e_{\rm p}\cos(\varpi_{\rm p}-\varpi)
    }{
    \frac14\alpha b_{3/2}^{(1)}(\alpha_0)-(\Delta\omega_{\rm ext}/n)/\mu_{\rm p} 
    }.
\end{align}
While $\varpi_{\rm p}-\varpi$ has an equilibrium, there is a divergence near $\Delta\omega_{\rm ext}=\Delta\omega_{\rm crit,sec}$, where
\begin{equation}\label{eq:Domcritsec}
    \Delta\omega_{\rm crit,sec} \simeq \frac14\left(\frac{j}{j+1}\right)\alpha b_{3/2}^{(1)}(\alpha_0)\mu_{\rm p} n_{\rm p}.
\end{equation}
Here we have used the resonant relation $n\simeq jn_{\rm p}/(j+1)$.
This ``divergence'' is the result of the apsidal precession resonance, which occurs when the precession rate of the planet $\omega_{\rm p}$ matches the net precession rate of the test particle, $\omega+(1/4)\alpha b_{3/2}^{(1)}(\alpha)\mu_{\rm p} n$.
For $j=2$, we have $ \Delta\omega_{\rm crit,sec}\simeq 1.54\mu_{\rm p} n_{\rm p}$.
We see in Figure~\ref{fig:capgrid} and the middle panel of Figure~\ref{fig:ressectm} that this value fits the spike of disruption well.
Equations~(\ref{eq:eeq}) and (\ref{eq:Domcritsec}) also explain whether $\varpi_{\rm p}-\varpi$ librates around 0 or $\pi$.
Since $e_{\rm eq,sec}>0$, if $\Delta\omega_{\rm ext}>\Delta\omega_{\rm crit,sec}$, then the denominator is negative, and we must have $\varpi_{\rm p}-\varpi=\pi$ so that the numerator is negative.
Analogously, $\varpi_{\rm p}-\varpi=0$ for the case where $\Delta\omega_{\rm ext}<\Delta\omega_{\rm crit,sec}$, and the denominator is positive.

\section{Discussion}\label{sec:discussion}
The problem we have studied in the previous sections closely mirrors the capture of a TNO into an external resonance with Neptune, which has a mass ratio of roughly $\mu_{\rm p}=5\times10^{-5}$.
Smooth outward migration of Neptune is a promising theory for dynamically sculpting the Kuiper belt and populating its resonances~\citep{malhotraOriginPlutoOrbit1995}.
For the current locations and masses of Neptune, Uranus, Saturn, and Jupiter, the cumulative differential apsidal precession on Neptune and a test particle in the 2:3 MMR is
\begin{align}
    \Delta\omega_{\rm ext} \approx 6.85\times10^{-5}n_{\rm p} \equiv \Delta\omega_{\rm GP}.
\end{align}
In reality, Neptune's precession rate was higher during the migration, due to its closer proximity to the Sun.
Resonance overlap is likely negligible for Neptune's resonance sweeping since $\Delta\omega_{\rm GP}\ll\Delta\omega_{\rm crit}$.
However, the critical secular resonance frequency $\Delta\omega_{\rm crit,sec}$ is comparable to $\Delta\omega_{\rm GP}$.

Our results could be used to constrain Neptune’s orbital evolution history, but detailing all the implications is beyond the scope of this study.
Comparing $\Delta\omega_{\rm GP}$ with the results in Figure~\ref{fig:capgrid}, we see that differential apsidal precession could affect the resonance capture, depending on Neptune's eccentricity during capture.
Non-captured particles near the secular resonance, as in Figure~\ref{fig:disruptex}, have their eccentricities excited to orbit crossing with Neptune, which implies that this is a potential pathway to populate the scattered disk.
Models with smooth, low-eccentricity migration of Neptune often predict a resonant-to-non-resonant population ratio that is too high \citep[e.g.][]{hahnNeptunesMigrationStirredUp2005,nesvornyEvidenceSlowMigration2015}, something which our findings could help rectify.
However, other migration scenarios have been put forward, such as ``grainy'' migration or a ``jumping-Neptune'' model \citep[for a review, see][]{nesvornyDynamicalEvolutionEarly2018}.
Nevertheless, our results are likely still relevant since all scenarios must include the effects of differential apsidal precession induced by the inner giant planets.
Most of these studies use N-body simulations that already account for the effects of the giant planets, so our study could help explain and interpret their results.

\subsection{Migration Timescale}
The planet's migration timescale can affect the resonance capture outcomes.
In the third panel of Figure~\ref{fig:ressectm}, we consider the capture problem for $\mu_{\rm p}=5\times10^{-5}$ but for 10 times faster migration rate than in Figure~\ref{fig:capgrid} (left), $\tau_m=10^5P_{\rm p,0}$.
Capture is generally more secure for larger $e_{\rm p}$ except when $\Delta\omega_{\rm ext}$ is very close to $\Delta\omega_{\rm crit}$ or $\Delta\omega_{\rm crit,sec}$.
Fully diagnosing this observation is beyond the scope of this paper, but we speculate that there are two reasonable explanations.
One is that, because of the faster migration time, the particle spends less time in the chaotic regime when the resonances overlap around $\Delta\omega_{\rm crit}$. 
The magnitude of the resonance parameter $k$, which sets $\theta$'s width in frequency space, grows rapidly so that resonance overlap ends sooner.
The second is that the transition from $\varpi_{\rm p}-\varpi$ librating to circulating occurs more quickly, reducing the amplitude of eccentricity excitation near the secular resonance at $\Delta\omega_{\rm crit,sec}$.

\subsection{Chaos}
In the regions of Figure~\ref{fig:capgrid} where there is a mix of resonance capture and disruption around $\Delta\omega_{\rm crit,sec}$, we find that the capture outcome is sensitive to the initial conditions.
We ran two integrations with identical $\mu_{\rm p}$, $e_{\rm p}$, $\omega$, and $\omega_{\rm p}$, differing only in the initial $\varpi$: in one, resonance is disrupted, and in the other, resonance capture is secure.
This fact, considered together with the FLI mapping (Figure~\ref{fig:5e-5FLI}) and the visual characteristics of $\theta$ and $\theta_{\rm p}$ in Figure~\ref{fig:disruptoverlapex}, suggests that the disruption in MMRs under differential apsidal precession is chaotic in both the secular resonance regime and resonance overlap regime.

\subsection{Apsidal Alignment}
In \citet{launeApsidalAlignmentAntialignment2022}, we found that $\theta$ and the combined resonant angle (see Section~\ref{sec:combres}) determined the behavior of $\varpi_{\rm p}-\varpi$ during and after resonance capture.
In that study, we included eccentricity damping of the test particle due to its interaction with a disk.
This interaction forced the test particle to a finite equilibrium eccentricity, which in turn determined the behavior of $\varpi_{\rm p}-\varpi$.
By contrast, all capture scenarios in the present study result in apsidal circulation, because there is no eccentricity damping.

\subsection{Comparable Mass Planets}
Capture into MMRs among comparable-mass planets undergoing differential apsidal precession is of astrophysical interest, since it is observed that external giant planets are common around inner planetary (super-Earth) systems~\citep[e.g.][]{zhuSuperEarthColdJupiter2018,bryanExcessJupiterAnalogs2019,vanZandt2025AJ....169..235V}.
We investigated this scenario with a parameterized model for disk-driven eccentricity damping and migration, and we found that capture is only affected if the giant planet is very massive, close-in, or highly eccentric.
We attribute this to the gas disk-induced eccentricity damping of the migrating planets. 
\citet{murrayEffectsDiskInduced2022} studied the effect of disk-induced apsidal precession and found that MMR is always captured for comparable-mass planets.
They also found that precession can lead to shifts in their equilibrium eccentricities, the inner planet's eccentricity diverges due to damping if the inner planet migrates into a gap, and rapid disk evaporation can destabilize the resonance.

\section{Summary}\label{sec:conclusion}
The capture of a test particle by a planet into the first-order $j:j+1$ MMR due to convergent migration is a well-studied problem in celestial mechanics.
For a finite planet eccentricity $e_{\rm p}$, the two resonance angles $\theta$ and $\theta_{\rm p}$ (see equations~\ref{eq:resangles}~and~\ref{eq:resangles1}) can typically be combined via a series of canonical transformations so that the system becomes a single-degree-of-freedom Hamiltonian.
Here we examine the effect of a finite differential apsidal precession, $\Delta\omega_{\rm ext}$, induced by a generic external source.
This effectively splits the two sub-resonances associated with $\theta$ and $\theta_{\rm p}$, and the standard reduction fails.
Such a splitting gives rise to resonance overlap and the possibility of chaos and therefore can significantly influence the resonance capture process.

We started by investigating the dynamics of the $\theta$-resonance coupled to $\theta_{\rm p}$ without the secular forcing from the planet on the test particle and without planet migration.
We calculated the separatrices of the individual resonances as a model for determining when overlap could occur.
By plotting phase-space trajectories and the Fast Lyapunov Indicator (FLI) for various values of $\Delta\omega_{\rm ext}$, we found signatures of chaos when the separatrices cross.
Next, we made the planet migrate, included its secular effects on the test particle, and integrated a random sampling of initial systems for a range of $e_{\rm p}$ and $\Delta\omega_{\rm ext}$.
We identified two distinct channels for the disruption of resonance and found that capture remains secure when $\Delta\omega_{\rm ext}$ and/or $e_{\rm p}$ are small (see Figure~\ref{fig:capgrid}).
In one channel, both $\theta$ and $\theta_{\rm p}$ evolve together chaotically, ultimately leading to escape from the resonance.
We attribute this behavior to the resonance overlap phenomenon, which occurs when $\Delta\omega_{\rm ext}\simeq\Delta\omega_{\rm crit}$ (see equation~\ref{eq:Domcrit}).

In addition, we found another disruption mechanism linked to the apsidal angle $\varpi_{\rm p}-\varpi=\theta-\theta_{\rm p}$.
In this channel, the particle's eccentricity is excited during the MMR crossing, leading to the disruption of the system.
This eccentricity growth arises from the apsidal precession resonance, which occurs when $\Delta\omega_{\rm ext}$ matches the apsidal precession rate of the test particle driven by the planet's secular forcing, i.e. $\Delta\omega_{\rm ext}\simeq\Delta\omega_{\rm crit,sec}$ (see equation~\ref{eq:Domcritsec}).
As a result, MMR capture can fail even for very small values of $e_{\rm p} \approx 0.001$.
This process is also likely chaotic, as we found that the capture process is sensitive to the initial conditions and the boundary between capture and disruption is fuzzy.

``External'' apsidal precession can arise from many different sources---including massive disks, $J_2$ of the central star, and additional companions—making it a common feature when studying MMR capture.
Our results demonstrate that even a small $\Delta\omega_{\rm ext}$ can have a significant impact on the possibility and probability of resonance capture during convergent migration, as a result of resonance overlap and secular apsidal precession resonance.
In the context of resonant TNO capture, our findings shed light on the role of perturbations from the inner giant planets, and may help constrain the evolution of the early Solar System.

\bibliographystyle{apalike}
\bibliography{main}

\begin{thebibliography}{}

\bibitem[Bryan et~al., 2019]{bryanExcessJupiterAnalogs2019}
Bryan, M.~L., Knutson, H.~A., Lee, E.~J., Fulton, B.~J., Batygin, K., Ngo, H., and Meshkat, T. (2019).
\newblock An {{Excess}} of {{Jupiter Analogs}} in {{Super-Earth Systems}}.
\newblock {\em The Astronomical Journal}, 157(2):52.

\bibitem[Deck et~al., 2013]{deck13_first_order_reson_overl_stabil}
Deck, K.~M., Payne, M., and Holman, M.~J. (2013).
\newblock First-order resonance overlap and the stability of close two-planet systems.
\newblock {\em The Astrophysical Journal}, 774(2):129.

\bibitem[El~Moutamid et~al., 2014]{elmoutamidCouplingCorotationLindblad2014}
El~Moutamid, M., Sicardy, B., and Renner, S. (2014).
\newblock Coupling between corotation and {{Lindblad}} resonances in the presence of secular precession rates.
\newblock {\em Celestial Mechanics and Dynamical Astronomy}, 118(3):235--252.

\bibitem[El~Moutamid et~al., 2017]{elmoutamidDerivationCaptureProbabilities2017}
El~Moutamid, M., Sicardy, B., and Renner, S. (2017).
\newblock Derivation of capture probabilities for the corotation eccentric mean motion resonances.
\newblock {\em Monthly Notices of the Royal Astronomical Society}, 469(2):2380--2386.

\bibitem[Fernandez and Ip, 1984]{fernandezDynamicalAspectsAccretion1984}
Fernandez, J.~A. and Ip, W.-H. (1984).
\newblock Some dynamical aspects of the accretion of {{Uranus}} and {{Neptune}}: {{The}} exchange of orbital angular momentum with planetesimals.
\newblock {\em Icar}, 58(1):109--120.

\bibitem[Gomes, 2000]{gomesPlanetaryMigrationPlutino2000}
Gomes, R.~S. (2000).
\newblock Planetary {{Migration}} and {{Plutino Orbital Inclinations}}.
\newblock {\em The Astronomical Journal}, 120:2695--2707.

\bibitem[{Gomes}, 2003]{gomes2003Icar..161..404G}
{Gomes}, R.~S. (2003).
\newblock {The origin of the Kuiper Belt high-inclination population}.
\newblock {\em \icarus}, 161(2):404--418.

\bibitem[{Hahn} and {Malhotra}, 2005]{hahnNeptunesMigrationStirredUp2005}
{Hahn}, J.~M. and {Malhotra}, R. (2005).
\newblock {Neptune's Migration into a Stirred-Up Kuiper Belt: A Detailed Comparison of Simulations to Observations}.
\newblock {\em \aj}, 130(5):2392--2414.

\bibitem[Henrard and Lemaitre, 1983]{henrardSecondFundamentalModel1983a}
Henrard, J. and Lemaitre, A. (1983).
\newblock A second fundamental model for resonance.
\newblock {\em Celestial Mechanics}, 30(2):197--218.

\bibitem[Henrard et~al., 1986]{henrardReducingTransformationApocentric1986}
Henrard, J., Lemaitre, A., Milani, A., and Murray, C.~D. (1986).
\newblock The reducing transformation and {{Apocentric Librators}}.
\newblock {\em Celestial mechanics}, 38(4):335--344.

\bibitem[Laune et~al., 2022]{launeApsidalAlignmentAntialignment2022}
Laune, J.~T., Rodet, L., and Lai, D. (2022).
\newblock Apsidal alignment and anti-alignment of planets in mean-motion resonance: Disc-driven migration and eccentricity driving.
\newblock {\em Monthly Notices of the Royal Astronomical Society}, 517(3):4472--4488.

\bibitem[{Malhotra}, 1993]{malhotraOriginPlutosPeculiar1993}
{Malhotra}, R. (1993).
\newblock {The origin of Pluto's peculiar orbit}.
\newblock {\em \nat}, 365(6449):819--821.

\bibitem[Malhotra, 1995]{malhotraOriginPlutoOrbit1995}
Malhotra, R. (1995).
\newblock The {{Origin}} of {{Pluto}}'s {{Orbit}}: {{Implications}} for the {{Solar System Beyond Neptune}}.
\newblock {\em AJ}, 110:420.

\bibitem[Murray and Dermott, 2000]{murraySolarSystemDynamics2000}
Murray, C.~D. and Dermott, S.~F. (2000).
\newblock {\em Solar {{System Dynamics}}}.
\newblock Cambridge University Press, Cambridge.

\bibitem[Murray et~al., 2022]{murrayEffectsDiskInduced2022}
Murray, Z., Hadden, S., and Holman, M.~J. (2022).
\newblock The {{Effects}} of {{Disk Induced Apsidal Precession}} on {{Planets Captured}} into {{Mean Motion Resonance}}.
\newblock {\em The Astrophysical Journal}, 931(1):66.

\bibitem[{Murray-Clay} and Schlichting, 2011]{murray-clayUSINGKUIPERBELT2011}
{Murray-Clay}, R.~A. and Schlichting, H.~E. (2011).
\newblock {{USING KUIPER BELT BINARIES TO CONSTRAIN NEPTUNE}}'{{S MIGRATION HISTORY}}.
\newblock {\em The Astrophysical Journal}, 730(2):132.

\bibitem[{Nesvorn{\'y}}, 2015]{nesvornyEvidenceSlowMigration2015}
{Nesvorn{\'y}}, D. (2015).
\newblock {Evidence for Slow Migration of Neptune from the Inclination Distribution of Kuiper Belt Objects}.
\newblock {\em \aj}, 150(3):73.

\bibitem[{Nesvorn{\'y}}, 2018]{nesvornyDynamicalEvolutionEarly2018}
{Nesvorn{\'y}}, D. (2018).
\newblock {Dynamical Evolution of the Early Solar System}.
\newblock {\em \araa}, 56:137--174.

\bibitem[{Peale}, 1976]{peale1976ARA&A..14..215P}
{Peale}, S.~J. (1976).
\newblock {Orbital resonance in the solar system.}
\newblock {\em \araa}, 14:215--246.

\bibitem[Petit et~al., 2020]{petitResonanceK219System2020}
Petit, A.~C., Petigura, E.~A., Davies, M.~B., and Johansen, A. (2020).
\newblock Resonance in the {{K2-19}} system is at odds with its high reported eccentricities.
\newblock {\em arXiv:2003.04931 [astro-ph]}.

\bibitem[Skokos et~al., 2016]{skokosChaosDetectionPredictability2016}
Skokos, C., Gottwald, G.~A., and Laskar, J., editors (2016).
\newblock {\em Chaos {{Detection}} and {{Predictability}}}, volume 915 of {\em Lecture {{Notes}} in {{Physics}}}.
\newblock Springer, Berlin, Heidelberg.

\bibitem[{Van Zandt} et~al., 2025]{vanZandt2025AJ....169..235V}
{Van Zandt}, J., {Petigura}, E.~A., {Lubin}, J., {Weiss}, L.~M., {Turtelboom}, E.~V., {Fetherolf}, T., {Murphy}, J. M.~A., {Crossfield}, I. J.~M., {Gilbert}, G.~J., {Mo{\v{c}}nik}, T., {Batalha}, N.~M., {Dressing}, C., {Fulton}, B., {Howard}, A.~W., {Huber}, D., {Isaacson}, H., {Kane}, S.~R., {Robertson}, P., {Roy}, A., {Angelo}, I., {Behmard}, A., {Beard}, C., {Chontos}, A., {Dai}, F., {Giacalone}, S., {Hill}, M.~L., {Holcomb}, R., {Howell}, S.~B., {Mayo}, A.~W., {Pidhorodetska}, D., {Polanski}, A.~S., {Rogers}, J., {Rosenthal}, L.~J., {Rubenzahl}, R.~A., {Scarsdale}, N., {Tyler}, D., {Yee}, S.~W., and {Zink}, J. (2025).
\newblock {The TESS{\textendash}Keck Survey. XXIV. Outer Giants May Be More Prevalent in the Presence of Inner Small Planets}.
\newblock {\em \aj}, 169(5):235.

\bibitem[Wisdom, 1986]{wisdomCanonicalSolutionTwo1986}
Wisdom, J. (1986).
\newblock Canonical solution of the two critical argument problem.
\newblock {\em Celestial Mechanics}, 38:175--180.

\bibitem[Zhu and Wu, 2018]{zhuSuperEarthColdJupiter2018}
Zhu, W. and Wu, Y. (2018).
\newblock The {{Super Earth-Cold Jupiter Relations}}.
\newblock {\em The Astronomical Journal}, 156:92.

\end{thebibliography}

\clearpage
\appendix
\section{Variational Equations for the Fast Lyapunov Indicator Method}\label{app:vareqs}
We scale the problem using $n_{\rm p}$ and $a_{\rm p}$, and we denote $n_{\rm p}t=\tau$.
We follow the method in \citet{skokosChaosDetectionPredictability2016} to compute the FLI.
We begin by transforming the time-dependent Hamiltonian~(\ref{eq:H}) into an autonomous form.
We accomplish this via a time-dependent generating function,
\begin{align}
    S(\Omega,\Theta,-\varpi,\lambda) = &\Omega \left( \varpi_{\rm p}+\omega_{\rm p}\tau-\varpi   \right)\nonumber\\
    &+ \Theta \left((j+1) \lambda + \gamma - j \tau\right),
\end{align}
so that the ``old'' canonical variables are related to the ``new'' ones by
\begin{align}
    \Gamma &= \Omega + \Theta \\
    \Lambda &= (j+1) \Theta \\
    -\varpi&=\Delta\varpi -( \varpi_{\rm p}+ (\omega_{\rm p}/n_{\rm p})\tau )\\
    \lambda &=\frac{- \Delta\varpi + j \tau + (\omega_{\rm p}/n_{\rm p}) \tau + \theta + \varpi_{\rm p}}{(j+1)}. 
\end{align}
The new Hamiltonian is
\begin{align}\label{eq:HFLI}
    H'=- &\frac{1}{2 \Theta^{2} \left(j + 1\right)^{2}}  \\
    &\Omega (\omega_{\rm p}/n_{\rm p}) - j\Theta  - (\omega/n_{\rm p}) \left(\Omega + \Theta\right) \nonumber\\
    &- \frac{ \beta \mu_{\rm p} \sqrt{2(\Omega + \Theta)} \cos{\left(\theta \right)}}{\sqrt{(j + 1)\Theta}}\\
    &- \beta_{\rm p} e_{\rm p} \mu_{\rm p} \cos{\left(\Delta\varpi - \theta \right)}, 
\end{align}
where we have added $\partial S/\partial \tau$.

We compute the FLI for the vector of new canonical variables, $\mathbf x \equiv (\theta,\Delta\varpi,\Theta,\Omega)$, and define the phase space vector $\mathbf y=(y_1,\ldots,y_4)$ as 
\begin{equation}
    \mathbf y\equiv\frac{d\mathbf x}{dt},
\end{equation}
where $d\mathbf x/d\tau$ is computed via Hamilton's equations for the Hamiltonian~(\ref{eq:HFLI}).
Then we derive the full variational equations as
\begin{equation}
    \dot{\mathbf y}\equiv\frac{d\mathbf y}{d\tau} = \frac{d\mathbf y}{d\mathbf x}\frac{d\mathbf x}{d\tau}=\frac{d\mathbf y}{d\mathbf x}\mathbf y,
\end{equation}
where $d\mathbf y/d\mathbf x$ is a $4\times 4$ matrix.
Its components are given as follows:
\begin{align}
    \dot y_1 =&~ y_{1} \left(\frac{\sqrt{2} \beta \mu_{\rm p} \sin{\left(\theta \right)}}{2 \sqrt{\Theta} \sqrt{\Omega + \Theta} \sqrt{j + 1}} 
    - \frac{\sqrt{2} \beta \mu_{\rm p} \sqrt{\Omega + \Theta} \sin{\left(\theta \right)}}{2 \Theta^{\frac{3}{2}} \sqrt{j + 1}}\right) \nonumber\\
    &+ y_{3} \left(- \frac{3 a_{\rm p}}{\Theta^{4} \left(j + 1\right)^{2}} + \frac{\sqrt{2} \beta \mu_{\rm p} \cos{\left(\theta \right)}}{4 \sqrt{\Theta} \left(\Omega + \Theta\right)^{\frac{3}{2}} \sqrt{j + 1}} 
    + \frac{\sqrt{2} \beta \mu_{\rm p} \cos{\left(\theta \right)}}{2 \Theta^{\frac{3}{2}} \sqrt{\Omega + \Theta} \sqrt{j + 1}} 
    - \frac{3 \sqrt{2} \beta \mu_{\rm p} \sqrt{\Omega + \Theta} \cos{\left(\theta \right)}}{4 \Theta^{\frac{5}{2}} \sqrt{j + 1}}\right) \nonumber\\
    &+ y_{4} \left(\frac{\sqrt{2} \beta \mu_{\rm p} \cos{\left(\theta \right)}}{4 \sqrt{\Theta} \left(\Omega + \Theta\right)^{\frac{3}{2}} \sqrt{j + 1}}
    + \frac{\sqrt{2} \beta \mu_{\rm p} \cos{\left(\theta \right)}}{4 \Theta^{\frac{3}{2}} \sqrt{\Omega + \Theta} \sqrt{j + 1}}\right),
\end{align}
\begin{align}
    \dot y_2 = y_{3} \left(\frac{\sqrt{2} \beta \mu_{\rm p} \cos{\left(\theta \right)}}{4 \sqrt{\Theta} \left(\Omega + \Theta\right)^{\frac{3}{2}} \sqrt{j + 1}}
    + \frac{\sqrt{2} \beta \mu_{\rm p} \cos{\left(\theta \right)}}{4 \Theta^{\frac{3}{2}} \sqrt{\Omega + \Theta} \sqrt{j + 1}}\right) + \frac{\sqrt{2} \beta \mu_{\rm p} y_{1} \sin{\left(\theta \right)}}{2 \sqrt{\Theta} \sqrt{\Omega + \Theta} \sqrt{j + 1}}
    + \frac{\sqrt{2} \beta \mu_{\rm p} y_{4} \cos{\left(\theta \right)}}{4 \sqrt{\Theta} \left(\Omega + \Theta\right)^{\frac{3}{2}} \sqrt{j + 1}},
\end{align}
\begin{align}
    \dot y_3 =&\beta_{\rm p} e_{\rm p} \mu_{\rm p} y_{2} \cos{\left(\Delta\varpi - \theta \right)}
    + y_{1} \left(- \beta_{\rm p} e_{\rm p} \mu_{\rm p} \cos{\left(\Delta\varpi - \theta \right)}
    - \frac{\sqrt{2} \beta \mu_{\rm p} \sqrt{\Omega + \Theta} \cos{\left(\theta \right)}}{\sqrt{\Theta} \sqrt{j + 1}}\right) \nonumber\\
    &+ y_{3} \left(- \frac{\sqrt{2} \beta \mu_{\rm p} \sin{\left(\theta \right)}}{2 \sqrt{\Theta} \sqrt{\Omega + \Theta} \sqrt{j + 1}} 
    + \frac{\sqrt{2} \beta \mu_{\rm p} \sqrt{\Omega + \Theta} \sin{\left(\theta \right)}}{2 \Theta^{\frac{3}{2}} \sqrt{j + 1}}\right) - \frac{\sqrt{2} \beta \mu_{\rm p} y_{4} \sin{\left(\theta \right)}}{2 \sqrt{\Theta} \sqrt{\Omega + \Theta} \sqrt{j + 1}},
\end{align}
\begin{align}
    &\dot y_4=\beta_{\rm p} e_{\rm p} \mu_{\rm p} y_{1} \cos{\left(\Delta\varpi - \theta \right)} - \beta_{\rm p} e_{\rm p} \mu_{\rm p} y_{2} \cos{\left(\Delta\varpi - \theta \right)}.
\end{align}
We integrate the entire system as
\begin{equation}
    \frac{d}{d\tau}\begin{bmatrix}\mathbf x\\\mathbf y\end{bmatrix} =\begin{bmatrix}\mathbf y\\ \dot{\mathbf y} \end{bmatrix}
    = \begin{bmatrix}\mathbf I \\ \frac{d\mathbf y}{d\mathbf x}\end{bmatrix}\begin{bmatrix}\mathbf y\end{bmatrix}.
\end{equation}
We initialize $\mathbf x_0$ from the initial orbit specified in osculating orbital elements.
For the initial conditions of $\mathbf y$, we repeat the integration four times for the following initial conditions:
\begin{equation}
    \mathbf y^i_0\in\left\{\begin{bmatrix}
        1 \\
        0 \\
        0 \\
        0 \\
    \end{bmatrix},
    \begin{bmatrix}
        0 \\
        1 \\
        0 \\
        0 \\
    \end{bmatrix},
    \begin{bmatrix}
        0 \\
        0 \\
        1 \\
        0 \\
    \end{bmatrix},
    \begin{bmatrix}
        0 \\
        0 \\
        0 \\
        1 \\
    \end{bmatrix},\right\}
\end{equation}
where the label $i$ denotes the index of the nonzero entry.
We integrate to a stopping time of $\tau_{\rm f}=2\times10^3$ and find that longer integrations do not change our qualitative results.
We calculate the individual FLIs, FLI$^i$, for each of these initial conditions $\mathbf y_0^i$ as
\begin{equation}
    \mathrm{FLI}^i \equiv \mathrm{FLI}(\mathbf x_0,\mathbf y_0^i )=\log_{10}\|\mathbf y(\tau_{\rm f})\|.
\end{equation}
Note there is no normalization by $\mathbf y_0^i$ since $\mathbf \|\mathbf y_0^i\|=1$.
Then we compute the average to get the final FLI,
\begin{equation}
    \mathrm{FLI} = \frac14\sum_{i=1}^4 \mathrm{FLI}^i.
\end{equation}

\section{The $\theta$-Resonance Reduced Hamiltonian}
\label{app:thH}
Here, we isolate the $\cos\theta$ term in Hamiltonian~(\ref{eq:H}) by setting $\beta_{\rm p}=0$.
We aim to reduce the Hamiltonian to a single degree of freedom in $\theta$.
Let $n_c$ and $a_c$ be defined as in equation~(\ref{eq:nc}).
Suppose $a$ is near $a_{\rm c}$.
This implies $\Gamma\simeq\frac12\sqrt{GMa_{\rm c}}~e^2$.

First, we transform to the rotating frame with the generating function
\begin{equation}
    S_{\rm r} = \Gamma\omega t -\Gamma\varpi,
\end{equation}
which gives us the canonical pair $\Gamma$, $-\varpi_{\rm r}=-\varpi+\omega t$.
Next, we perform two successive canonical transformations and reduce $H$ to a single degree of freedom, $\theta$.
We label the transformed canonical coordinates and momenta as $(\lambda',-\varpi_{\rm r},\Lambda',\Gamma)$ and $(\lambda'',\theta,K,\Gamma)$, respectively (i.e., the transformations on $\Gamma$ are the identity).
The transformations are given by the type 2 generating functions
\begin{align}
    S' &= \Lambda'\left(\lambda-\frac{j\lambda_{\rm p}}{j+1}-\frac{\omega t}{j+1}\right)+\sqrt{GMa_{\rm c}}\lambda \\
    S'' &= \Gamma((j+1)\lambda'-\varpi_{\rm r})+K\lambda'.
\end{align}
Note that $S'$ is explicitly time-dependent through $\lambda_{\rm p}$.
The final coordinate is just $\theta$, conjugate to $\Gamma$.
The momentum $\Lambda$ transforms into
\begin{align}
    K = \Lambda - \sqrt{GMa_{\rm c}} - (j+1)\Gamma \equiv \frac12\sqrt{GMa_{\rm c}}k,
\end{align}
which defines $k$.
If we define $\delta a$ by $\delta a = a-a_{\rm c}$, $k$ becomes
\begin{align}
    k\simeq \frac{\delta a}{a_{\rm c}} - (j+1) e^2.
\end{align}
Note that $k$ is conserved because $\lambda''$ is not present in the new Hamiltonian.
If we now rescale the momentum variable by $\frac12\sqrt{GMa_c}$ and time by $n_{\rm p}$, we get the Hamiltonian~(\ref{eq:calH}) with the canonically conjugate pair $e^2\longleftrightarrow\theta$.

The critical value $k=k_{\rm sep}$ (equation~\ref{eq:ksep}), at which a separatrix appears, is determined by the condition that the cubic equation 
\begin{align}
  \left.e\dot\theta\right|_{\theta=0} = \left.e\frac{\partial\mathcal H/}{\partial e^2}\right|_{\theta=0}=0  
\end{align}
has a double root at $e=e_{\rm crit}$.
This corresponds to an unstable fixed point at $(e,\theta)=(e_{\rm crit},\theta)$, which the separatrix passes through.
Since $\omega/n_{\rm p}\ll 1$, $\omega$ has a negligible effect on $k_{\rm sep}$ and is omitted.
The primary influence $\omega$ has on the resonance is setting the location of $a_{\rm c}$.
Now we may utilize $k_{\rm sep}$ to estimate the SMA width (equivalently frequency/mean motion width) of the separatrix when it first appears, $\delta a_{\rm sep}$.
At $k=k_{\rm sep}$,
\begin{align}
    e^2_{\rm crit} = -\frac{k_{\rm sep}}{3(j+1)}.
\end{align}
We approximate the minimum $e$ along the separatrix as $e_{\rm min}=e_{\rm crit}$, where it crosses $\theta=0$.
We approximate the maximum as $e_{\rm max}=3e_{\rm crit}$, where the separatrix crosses $\theta=\pi$.
Since $k$ is conserved,
\begin{align}
    \frac{\delta a_{\rm max}}{a_c} - (j+1)e_{\rm max}^2 = \frac{\delta a_{\rm min}}{a_c} - (j+1) e_{\rm min}^2,
\end{align}
where $\delta a_{\rm max}$ and $\delta a_{\rm min}$ correspond to $\delta a$ at $e=e_{\rm max}$  and $e=e_{\rm min}$, respectively.
Substituting in $k_{\rm sep}$ for $e_{\rm crit}$, and defining $\delta a_{\rm sep}=\delta a_{\rm max}-\delta a_{\rm min}$, we arrive at equation~(\ref{eq:dasep}).

\section{Canonical Transformations of the Hamiltonian: 2 Degrees of Freedom}\label{app:canontrans}

To derive equation~(\ref{eq:Hbar}), we start from equation~(\ref{eq:H}), scale time by $n_{\rm p}$ and momentum by $\sqrt{GMa_{\rm p}}$, and perform four canonical transformations.
The transformations are performed with the simplifying assumption that
$\Gamma\approx\frac{1}{2\sqrt{\alpha_0}} e^2$.
This approximation holds because the dominant contribution to the resonant terms comes from variations in $e$.
The five different sets of canonically conjugate momenta and coordinates are labeled as
\begin{align}
    &\Lambda\longleftrightarrow\lambda,\quad\Gamma\longleftrightarrow\gamma,\nonumber\\
    &\Theta\longleftrightarrow\theta_{\rm p},\quad\Gamma'\longleftrightarrow\gamma',\nonumber\\
    &\Theta\longleftrightarrow\theta_{\rm p},\quad Y_1\longleftrightarrow X_1,\\
    &\Theta\longleftrightarrow\theta_{\rm p},\quad Y_2\longleftrightarrow X_2\nonumber,\\
    &\Theta\longleftrightarrow\theta_{\rm p},\quad \Phi\longleftrightarrow\phi.\nonumber
\end{align}
The set is expressed in equations (\ref{eq:Theta}) and (\ref{eq:phi}).
The rest are given by the following expressions:
\begin{align}
    &\Gamma' = \Gamma,\nonumber\\
    &\gamma'=\gamma, \nonumber\\
    &Y_1=\sqrt{2\Gamma'}\cos\gamma',\nonumber\\
    &X_1=\sqrt{2\Gamma'}\sin\gamma',\nonumber\\
    &Y_2=Y_1-\frac{\beta_{\rm p}e_{\rm p}}{\alpha^\frac14\beta},\\
    &X_2=X_1,\nonumber\\ 
    &\Phi = \frac12(X_2^2 + Y_2^2),\nonumber\\
    &\phi = \tan^{-1}\frac{X_2}{Y_2}\nonumber.
\end{align}
The type 2 generating functions for the first three transformations are
\begin{align}
    F_1(\Theta,\Gamma',\lambda,\gamma) 
    &=\Theta \left(\lambda- \frac{(\omega_{\rm{p,ext}}/n_{\rm p}) t}{j + 1} - \frac{jt}{j + 1}\right) \nonumber\\&+ \Gamma' \left((\omega_{\rm{p,ext}}/n_{\rm p}) t + \gamma\right),\\
    F_2(\Theta,Y_1,\theta_{\rm p},\gamma')
    &=\frac12Y_1^2\tan\gamma'+\Theta\theta_{\rm p},\\
    F_3(\Theta,Y_2,\theta_{\rm p},X_1)
    &= X_{1} \left(Y_{2} - \frac{\beta_{\rm p} e_{\rm p}}{\sqrt[4]{\alpha} \beta}\right)+\Theta\theta_{\rm p}.
\end{align}
The last transformation, to $\Phi$ and $\phi$, is generated by the type 3 generating function
\begin{align}
    F_4(\Theta,Y_2,\theta_{\rm p},\phi) = -\frac12 Y_2^2\tan\phi-\Theta\theta_{\rm p}.
\end{align}
The Hamiltonian is now
\begin{align}\label{eq:Hbar_unexpanded}
    \overline{\mathcal{H}} =& 
    -\frac{1}{2\Theta^2} -\left(\frac{j+\omega_{\rm p}/n_{\rm p}}{j+1}\right)\Theta+\Delta\omega_{\rm ext}\Phi\nonumber\\
    &-\sqrt{2\Phi}\alpha_0^{1/4} \beta\mu_{\rm p}\cos((j+1)\theta_{\rm p}+\phi)\nonumber\\
    &+\sqrt{2\Phi}\frac{\beta_{\rm p}}{\beta\alpha_0^{1/4}}e_{\rm p}\Delta\omega_{\rm ext}\cos\phi.
\end{align}
Next we define
\begin{align}
    \Theta_0=\left(\frac{j+1}{j+\omega_{\rm p}/n_{\rm p}}\right)^{1/3}
\end{align}
and use the generating function
\begin{align}
    F_5(\delta\Theta,\Phi,\theta_{\rm p},\phi)=(\delta\Theta+\Theta_0)\theta_{\rm p}+\Phi\phi
\end{align}
so that $\theta$, $\phi$, and $\Phi$ all remain unchanged, but the new momentum $\delta\Theta$ conjugate to $\theta_{\rm p}$ is 
\begin{align}
    \delta\Theta = \Theta-\Theta_0.
\end{align}
Assuming $\delta\Theta$ is small, we expand the first term in a Taylor series and arrive at the final form of the resonance Hamiltonian given in equation~(\ref{eq:Hbar}).
The terms linear in $\delta\Theta$ cancel out.
    
\end{document}